\documentclass[final,5p,times,twocolumn]{elsarticle}
\usepackage{amssymb}
\usepackage{amsmath}
\usepackage{graphicx}
\usepackage{subcaption}
\usepackage{hyperref}
\usepackage{algorithm}      
\usepackage{algorithmicx}   
\usepackage{algpseudocode}
\usepackage{multirow}
\usepackage{graphicx}
\usepackage{diagbox}
\newcommand{\tabincell}[2]{\begin{tabular}{@{}#1@{}}#2\end{tabular}}

\journal{Nuclear Physics B}

\begin{document}

\begin{frontmatter}

\title{A Deep Learning Framework for Identifying Weakly Chaotic, Strongly Chaotic, Resonant and Non-resonant Orbits in the Generalized Kicked Rotator} 

\author[a]{Jingyue Hao}
\author[a]{Zhiguo Xu}
\affiliation[a]{organization={School of Mathematics, Jilin University},
	city={Changchun 130012},country={P.R. China}}
\author[b]{Jian Zu\corref{cor1}}
\affiliation[b]{organization={Center for Mathematics and Interdisciplinary Sciences, School of Mathematics and Statistics, Northeast Normal University},
	city={Changchun 130024},country={P.R. China}}
\cortext[cor1]{Corresponding author: zuj100@nenu.edu.cn} 

\begin{abstract}
Identifying the types of orbits is an important topic in the study of chaotic dynamical systems.  Beyond the well-known distinctly chaotic and regular motions, we focus on dynamics occurring in regions where regular and chaotic motions coexist and intertwine, which potentially indicating weakly chaotic orbits. This intermediate regime lies between strongly chaotic dynamics, characterized by exponential sensitivity and completely non-chaotic, purely regular behavior. 
In this paper, we introduce a deep learning framework to identify the types of orbits in the  generalized kicked rotator system, which is challenging to study due to its complex and mixed chaotic behaviors. Our deep learning framework can be divided into two steps. First, we propose a novel algorithm that integrates the weighted Birkhoff average, the Lyapunov exponent, and the correlation dimension to identify weakly chaotic orbits. The algorithm categorizes orbits into four types: weakly chaotic, strongly chaotic, and regular orbits (which are further subdivided into resonant and non-resonant orbits), thereby creating a valuable dataset required for deep learning models. Second, we demonstrate that a well-trained 2D-CNN achieves high performance in accurately classifying orbits, largely because it effectively leverages the 2D structural information of the phase space relation. To our knowledge, this is the first paper to identify weakly chaotic orbits using deep learning methods. The method can be easily extend to other models.

\end{abstract}


\end{frontmatter}


\section{Introduction}
\label{sec1}

In the exploration of nonlinear dynamic systems, chaos theory, as one of the major scientific revolutions of the 20th century, has profoundly reshaped our understanding of complexity and predictability in nature.
It reveals that even deterministic systems can exhibit extreme sensitivity to initial conditions and long-term behavioral unpredictability-phenomena that are ubiquitous from celestial mechanics to ecological models. Within this framework, identifying different types of orbits becomes a critical endeavor, as it allows researchers to decipher the fundamental transitions between ordered and chaotic motions, thereby uncovering the underlying mechanisms governing system evolution. One illustrative paradigm for studying such dynamics is provided by two-dimensional area-preserving maps. In integrable systems, motion is regular and confined to invariant tori, with each orbit characterized by a specific rotation number. However, this orderly structure is highly fragile under perturbations. According to the Kolmogorov-Arnold-Moser (KAM) theory \cite{de2001tutorial}, when a small disturbance is introduced, most invariant tori persist, while others degenerate into isolated periodic orbits, resonant islands, or chaotic regions---giving rise to intricately interwoven dynamical patterns. The coexistence and interaction of these distinct orbital types not only exemplify the universal route from regularity to chaos but also serve as a cornerstone for understanding stability, controllability, and complexity in physical, biological, and engineering systems. 

Lyapunov exponent is a widely recognized method for identifying different dynamical phenomena, as it measures  the separation of nearby orbits. As noted by Eckmann and Ruelle \cite{eckmann1985ergodic} and Ott \cite{ott2002chaos}, the magnitude of Lyapunov exponents reflects the ``strength'' of chaotic dynamics. Distinctly positive Lyapunov exponents typically characterize strongly chaotic orbits, while values approaching zero suggest weakly chaotic orbits---a behavior characterized by subexponential divergence. Regions where regular and chaotic dynamics coexist often contain weakly chaotic orbits. These orbits represent an intermediate regime between strongly chaotic dynamics and purely regular dynamics. Macroscopically, weakly chaotic orbits exhibit anomalous dynamical features, such as dynamical aging and anomalous diffusion.  Typical examples include maps with indifferent fixed points, polygonal billiards, and Hamiltonian systems featuring sticky islands in phase space \cite{klages2013weak}.

Besides Lyapunov exponent, there are many other methods used to classify dynamical behaviors. Kolmogorov-Sinai entropy (K-S entropy) quantifies the degree of disorder or unpredictability in a dynamical system \cite{sinai2009kolmogorov}. A K-S entropy of zero corresponds to a regular system, where the future state can be determined from the initial conditions. In contrast, a positive value indicates that the system exhibits chaotic behavior. Pesin's theorem establishes the fundamental relationship between K-S entropy and Lyapunov exponent, demonstrating that under appropriate conditions, K-S entropy equals the sum of all positive Lyapunov exponents \cite{dorfman1999introduction, pesin1977characteristic}.  The weighted Birkhoff average (WBA) allows for quick and accurate distinction between regular and chaotic orbits \cite{das2016measuring,meiss2025resonance}. For chaotic orbits, WBA typically converges slowly due to mixing and sensitivity to initial conditions. In contrast, for regular orbits, the WBA converges significantly faster, often exhibiting super-convergence. 

The traditional methods mentioned above typically require explicit knowledge of the system’s governing equations or substantial trajectory data, which limits their applicability in scenarios with only observational data or limited data. Deep Learning (DL) offers a promising solution for classifying motion types from time series data, which enables more comprehensive cartographic studies at reduced computational costs \cite{barrio2023deep, uzun2024deep, celletti2022classification, lee2020deep}. Furthermore, DL can predict the future behavior of chaotic systems directly from time series data \cite{pathak2018model, zhang2024deep}.
In \cite{barrio2023deep}, Barrio et al. established the efficacy of convolutional neural networks (CNNs) for the binary classification of dynamical system orbits, demonstrating that a properly designed CNN achieves high accuracy in distinguishing regular from chaotic regimes using finite-time orbits substantially shorter than those required by traditional Lyapunov methods. In \cite{lee2020deep}, Lee et al. focused on binary orbit classification, systematically comparing multilayer perceptron (MLP), CNN, and long short-term memory (LSTM) architectures for categorizing time series from the Logistic map and Lorenz system into regular or chaotic categories. In \cite{uzun2024deep}, Uzun et al. explored binary system classification, employing a deep learning approach that converts time series into graphic images and uses transfer learning models like SqueezeNet and ResNet to achieve high accuracy in distinguishing between different chaotic systems. The seminal work by Celletti et al. \cite{celletti2022classification} addressed a three-class orbit classification problem, providing a systematic categorization of three fundamental motion types—chaotic, rotational, and librational—in pendulum-like systems and the spin-orbit model, with the InceptionTime CNN architecture exhibiting superior performance.

The standard map in particular has been widely used in diverse scientific fields, such as particle dynamics in accelerators \cite{izraelev1980nearly}, comet dynamics \cite{petrosky1986chaos}, and the autoionization of molecular Rydberg states \cite{benvenuto1994chaotic}. Inspired by its extensive applications and theoretical importance in chaos theory, the generalized standard map (also known as generalized kicked rotator) was recently constructed in \cite{cetin2022generalization}. This generalized version is considered explanatory and more appropriate for modeling complex systems that cannot be adequately reduced to the original standard map as a first approximation. Given the increased complexity of this generalized kicked rotator (GKR) system, identifying weak chaos and classifying orbits within it using deep learning (DL) has emerged as a significant research endeavor. The GKR system is defined as:
\begin{equation}
	\begin{aligned}
		x_{n+1} &= x_n + y_{n+1}, \\
		y_{n+1} &= y_n + K \sum_{j=1}^{M} \sin(2\pi j x_n), \\
	\end{aligned} \qquad \mod 1,
	\label{1}
\end{equation}
where $x_n$ represents the angular position of the particle after the $n$-th iteration, and $y_n$ the corresponding momentum. The term $K\sum_{j=1}^{M} \sin(2\pi j x_n)$ refers to the generalized force, where $K$ is a positive dimensionless map parameter that controls the extent of nonlinearity of the system, $M$ is a positive integer. This seemingly simple setup gives rise to rich and complex dynamics, including strong chaos, weak chaos, and invariant tori, making it an ideal testing ground for classifying different types of motion. 

In this paper, we propose a novel algorithm for identifying weakly chaotic orbits and classifying orbital dynamics in the GKR system into four distinct categories: strongly chaotic, weakly chaotic, resonant, and non-resonant orbits. A key innovation of our approach lies in the synergistic integration of three complementary methods---the weighted Birkhoff average, Lyapunov exponent, and correlation dimension---which collectively enable a more robust and accurate characterization of weak chaos, a regime often challenging to delineate with single-metric methods. To generate high-quality labeled data for deep learning, we first evolve a set of orbits from given initial conditions over sufficiently long timescales, applying our algorithm to assign a definitive dynamical category to each orbit. These algorithmically generated labels, derived from long-term iterations, are then paired with corresponding shorter finite-length orbital segments to construct the training and validation datasets. This strategy of using extended trajectories for label generation and shorter subsequences as model inputs effectively balances label reliability with data diversity, forming a scalable and practical foundation for subsequent DL-based classification.

In the deep learning component, we employ the 2D-CNN method to fully leverage the informational richness of image-based representations. This image-based approach (unlike direct time-series processing) captures inherent spatial patterns in the trajectory data. By analyzing features in these trajectory images, it effectively identifies orbit types. When trained on a dataset of indexes paired with their corresponding orbits, the model achieves outstanding classification performance, with accuracy exceeding 99\%. This significantly outperforms the classification accuracy of models like MLP, InceptionTime CNN, and their Transformer-based hybrids (i.e., MLP-Transformer, InceptionTime CNN-Transformer). Furthermore, the 2D-CNN demonstrates considerable generalization capability, confirming its robustness in handling varied dynamical regimes. To the best of our knowledge, this represents the first study employing a machine learning approach to classify chaotic dynamical systems into four distinct categories—strongly chaotic, weakly chaotic, resonant and non-resonant orbits.

The paper is structured as follows. In Section \ref{sec2}, we detail the algorithms used to identify the weakly chaotic orbits and classify the orbits into the aforementioned four types. In Section \ref{sec3}, we compare the classification performance of several different DL models and demonstrate that the 2D-CNN model achieves exceptional classification results. Finally, in Section \ref{sec4}, we summarize the main findings of our study and discussing their implications.

\section{Orbits classification}
\label{sec2}

This section presents an algorithm to identify weakly chaotic orbits and, for the first time, classifies orbits in the GKR system into four distinct categories. For convenient, we take $\mathbf{x_{n}} = (x_n,y_n)^T$, then Eq.\eqref{1} can be rewritten as:
\[
\mathbf{x_{n+1}} = \mathbf{f} (\mathbf{x_n} ) = \left( 
\begin{array}{c}
	x_n + y_n + K \sum\limits_{j=1}^{M} \sin(2\pi j x_n) \\
	y_n + K \sum\limits_{j=1}^{M} \sin(2\pi j x_n)
\end{array}
\right) \mod 1.
\]
Clearly, $\mathbf{f}$ is a map from $\mathbb{T}^2$ to itself. When $M=1$, the system is called the classical
standard map, which was introduced by Chirikov \cite{chirikov1979universal}. For \( K = 0 \), the system is completely integrable and every orbit lies on an invariant torus. For \( K \ll 1 \), according to the KAM theory, most of invariant tori are preserved, and resonant ``islands'' emerge. These islands are surrounded by ``chaotic sea''. 

\begin{figure}[htbp]
	\centering
	\begin{subfigure}[b]{0.3\linewidth}
		\centering
		\includegraphics[width=1.1\linewidth]{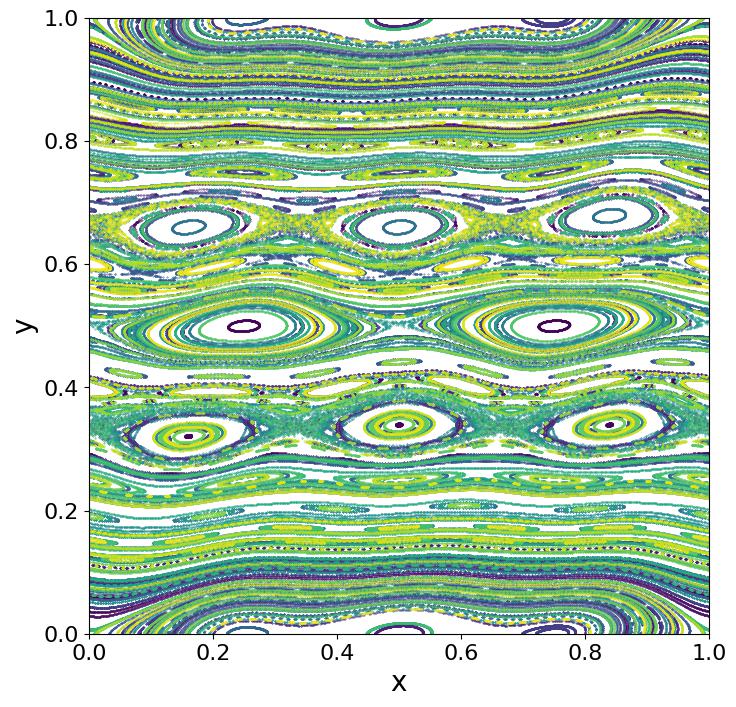}
		\caption{$K=0.01,M=3$}
		\label{fig1a}
	\end{subfigure}
	\hfill
	\begin{subfigure}[b]{0.3\linewidth}
		\centering
		\includegraphics[width=1.1\linewidth]{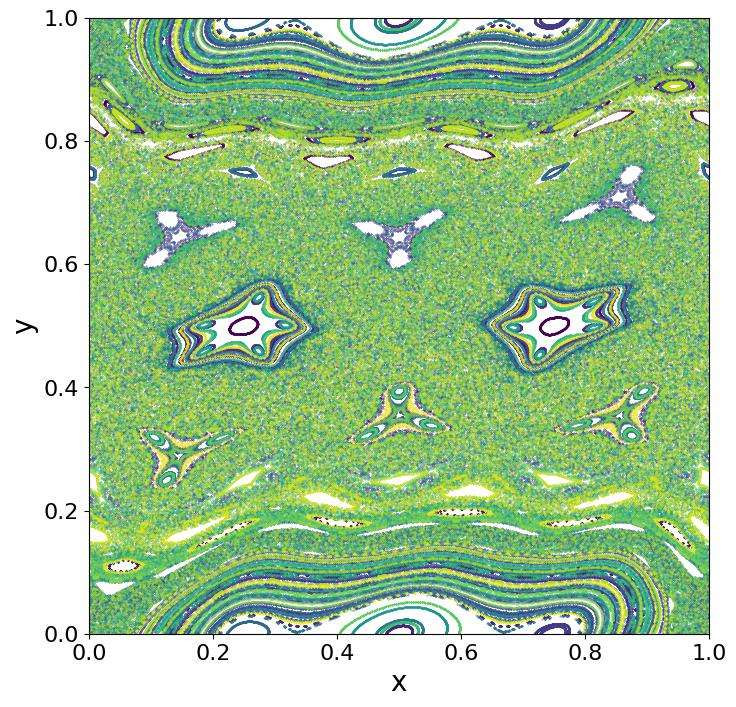}
		\caption{$K=0.03,M=3$}
		\label{fig1b}
	\end{subfigure}
	\hfill
	\begin{subfigure}[b]{0.3\linewidth}
		\centering
		\includegraphics[width=1.1\linewidth]{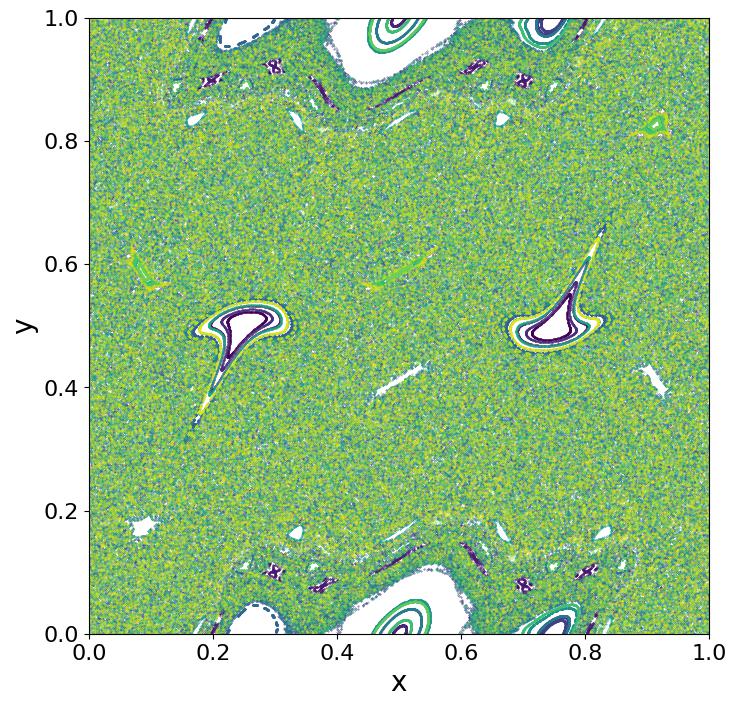}
		\caption{$K=0.1,M=3$}
		\label{fig1c}
	\end{subfigure}
	\vspace{1.0cm}
	\begin{subfigure}[b]{0.3\linewidth}
		\centering
		\includegraphics[width=1.1\linewidth]{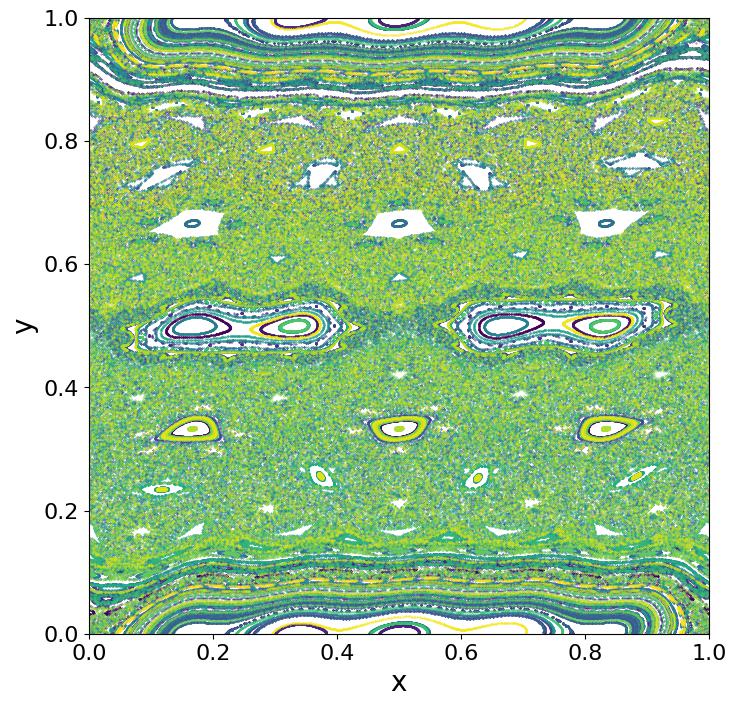}
		\caption{$K=0.01,M=5$}
		\label{fig1d}
	\end{subfigure}
	\hfill
	\begin{subfigure}[b]{0.3\linewidth}
		\centering
		\includegraphics[width=1.1\linewidth]{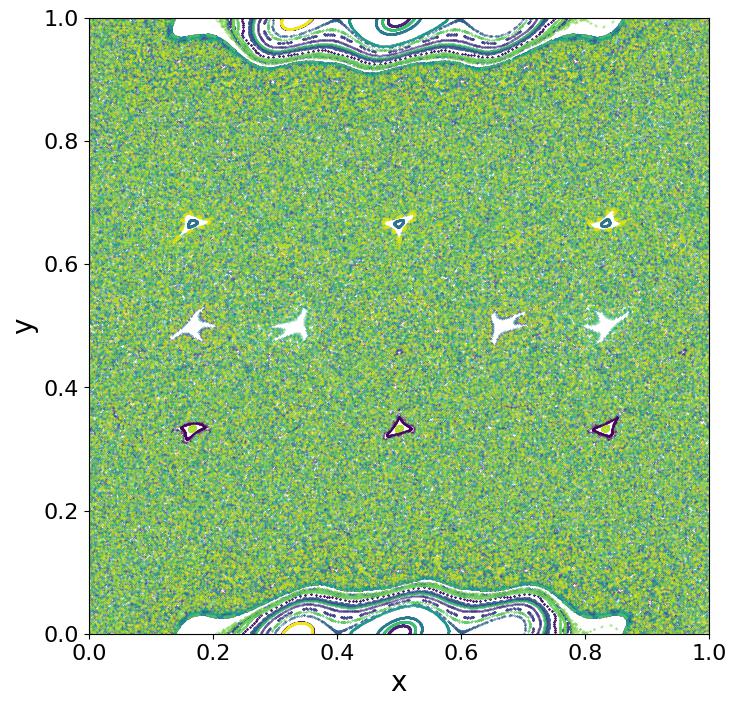}
		\caption{$K=0.03,M=5$}
		\label{fig1e}
	\end{subfigure}
	\hfill
	\begin{subfigure}[b]{0.3\linewidth}
		\centering
		\includegraphics[width=1.1\linewidth]{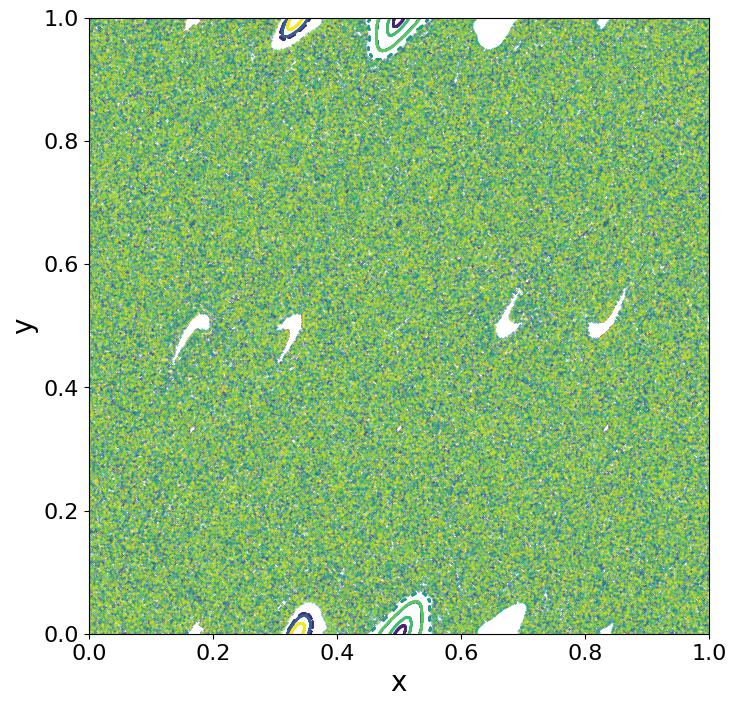}
		\caption{$K=0.1,M=5$}
		\label{fig1f}
	\end{subfigure}
	\caption{The orbits of the GKR system for $K=0.01,0.03,0.1$ with $M=3$ (Top) and $M=5$ (Bottom) from left to right. Here, we take $400$ different initial conditions, randomly distributed throughout the phase space and iterate each of these initial conditions through the generalized kicked rotator $1000$ steps. Different colors represent orbits with different initial conditions.}
	\label{fig1}
\end{figure}

Figure \ref{fig1} illustrates the phase space of the GKR system for varying parameters.  It is evident that as the parameters \( K \) and \( M \) increase, the dynamical behavior of the system undergoes significant changes: some invariant tori are preserved, while certain resonant islands gradually sink into the connected chaotic sea, as shown from left to right in both the top and bottom rows. This suggests the possible presence of weakly chaotic orbits during the transition from regularity to chaos.  

Figure \ref{fig1} illustrates the phase space of the GKR system for varying parameters.  It is evident that as the parameters \( K \) and \( M \) increase, the dynamical behavior of the system undergoes significant changes: some invariant tori are preserved, while certain resonant islands gradually sink into the connected chaotic sea, as shown from left to right in both the top and bottom rows. This suggests the possible presence of weakly chaotic orbits during the transition from regularity to chaos.  

\begin{figure}[htbp]
	\centering
	\begin{minipage}{0.48\textwidth}
		\centering
		\includegraphics[width=0.66\linewidth]{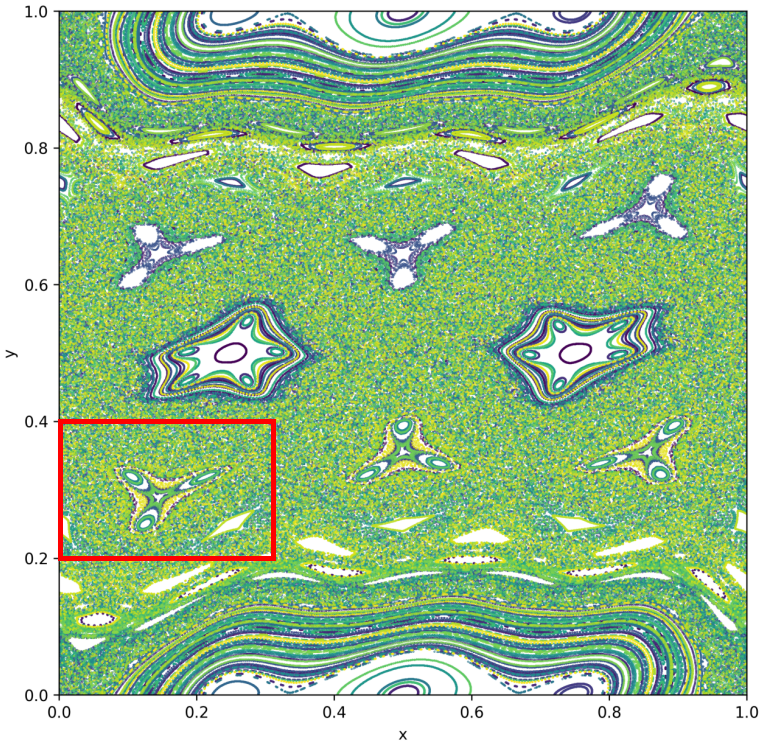}
		\caption*{(a) $K=0.03$, $M=3$}
		\label{fig2a}
	\end{minipage}
	\hfill
	\begin{minipage}{0.48\textwidth}
		\centering
		\begin{subfigure}{0.48\linewidth}
			\centering
			\includegraphics[width=4.0cm,height=2.67cm]{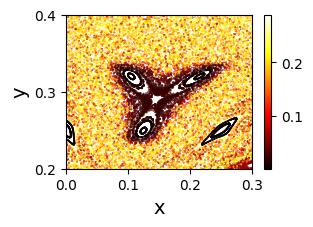}
			\subcaption{Lyapunov Exponents}
			\label{fig2b}
		\end{subfigure}
		\hfill
		\begin{subfigure}{0.48\linewidth}
			\centering
			\includegraphics[width=3.5cm,height=2.67cm]{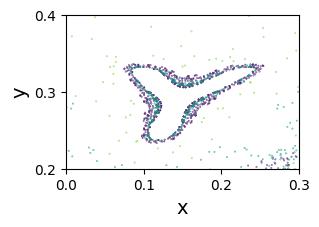}
			\subcaption{Weakly chaotic orbits}
			\label{fig2c}
		\end{subfigure}
	\end{minipage}
	\caption{(a) The orbits of the GKR system for $K=0.03$, $M=3$. (b) Lyapunov exponent heatmap  and (c) weakly chaotic orbits for the region $[0,0.3] \times [0.2,0.4]$ in (a).}
	\label{fig2}
\end{figure}

\begin{figure}[htbp]
	\centering
	\includegraphics[width=0.66\linewidth]{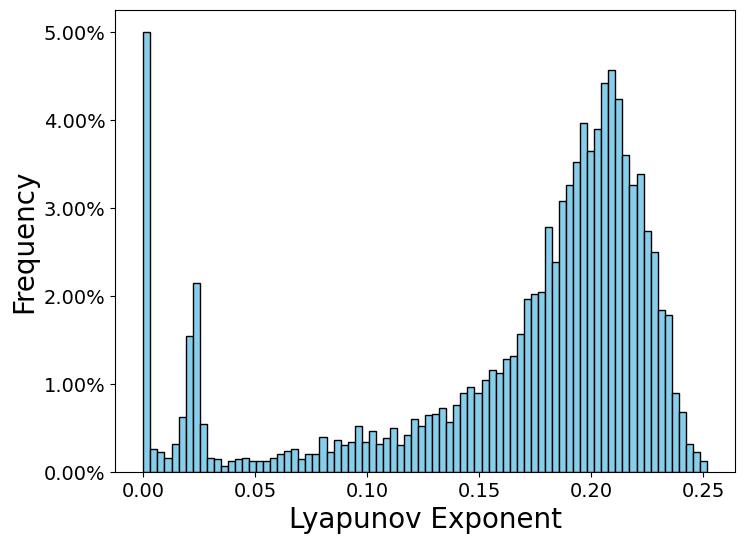}
	\caption{The histogram of Lyapunov exponents for the region $[0, 0.3] \times [0.2, 0.4]$ with the case $K=0.03, M=3$.}
	\label{fig3}
\end{figure}

As shown in Figure \ref{fig2}, resonant islands coexist with surrounding chaotic orbits within the region $[0.0,0.3] \times [0.2,0.4]$. The orbits around the edge of the resonant island are candidates for being weakly chaotic orbits. To quantitatively verify  this phenomenon, we calculate the Lyapunov exponents (finite-time approximations) within this region for the case $K = 0.03, M = 3$. Orbits lying on resonant islands exhibit relatively small Lyapunov exponents (on the order of $10^{-3}$). As orbits move away from these islands, the exponents increase continuously, reaching values around 0.25 in the central chaotic sea, which indicates strong chaos, see Figure \ref{fig2}b. If only regular orbits and the chaotic sea were present, a clear gap in the Lyapunov exponents would be expected. However, as illustrated in Figure \ref{fig3}, the exponents change gradually without a distinct separation. This continuous transition, in the absence of a clear gap, suggests the existence of weakly chaotic orbits forming an intermediate region between the resonant islands and the chaotic sea, as visualized in Figure \ref{fig2}c.

\begin{figure}[htbp]
	\centering
	\includegraphics[width=1.25\linewidth]{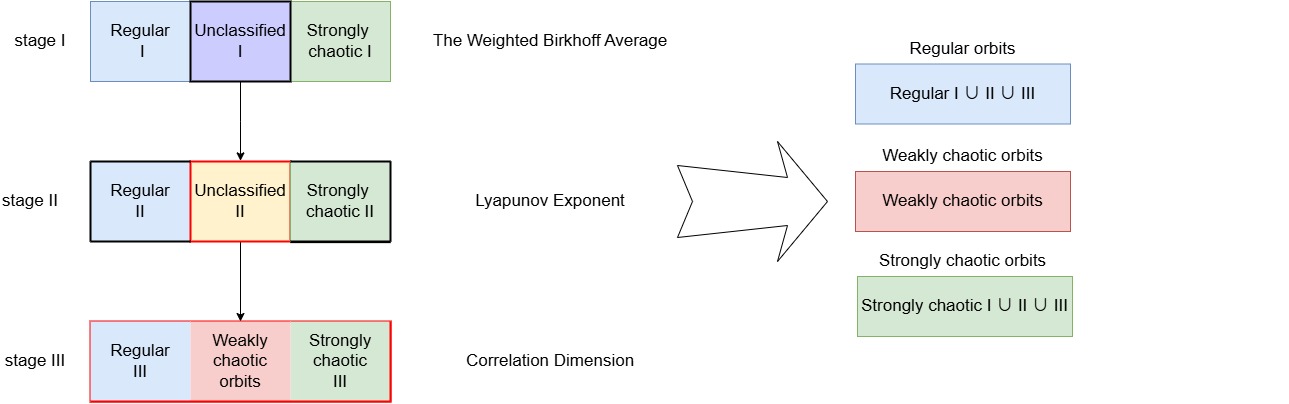}
	\caption{The framework outlines the classification process of different types of orbits in the Eq.\eqref{1}.}
	\label{fig4}
\end{figure}

We employ a hierarchical classification framework utilizing three methods sequentially to identify weakly chaotic orbits, as illustrated in Figure \ref{fig4}. First, the weighted Birkhoff average (WBA) method is used to identify an initial set of strongly chaotic orbits (Strongly chaotic I) and regular orbits (Regular I). The remaining orbits are labeled Unclassified I. Second, we compute the Lyapunov exponents (LE) for all orbits. Using the Regular I and Strongly chaotic I groups as references, we establish LE thresholds to classify a portion of the unclassified I orbits into strongly chaotic orbits (Strongly chaotic II) and regular orbits (Regular II). The orbits that remain unclassified after this step are labeled Unclassified II. Finally, we perform a correlation dimension (CD) analysis. By examining the CD distributions of the previously classified orbits (Strongly Chaotic I $\cup$ II and Regular I $\cup$ II), we establish two thresholds. Based on these thresholds, the Unclassified II orbits are classified into Strongly chaotic III, Regular III, and Weakly chaotic orbits. The detailed process is outlined below.

\subsection{The weighted Birkhoff average}
\label{sec2.1}

According to \cite{das2016quasiperiodicity}, the Birkhoff average of a function $h \in L^1(\mathbb{T}^2, \mathbb{R})$ along the trajectory of the map  $\mathbf{f}$ starting at $\mathbf{x_0}$ is defined by
\begin{equation}
	B_N(h)(\mathbf{x_0}) = \frac{1}{N} \sum_{n=0}^{N-1} h(\mathbf{f}^n(\mathbf{x_0})),
	\label{2}
\end{equation}
where $N$ is the number of iterations. Under mild hypotheses, the Birkhoff Ergodic Theorem (see Theorem 4.5.5. in \cite{brin2002introduction}) states that
\[
\lim_{N \to \infty} B_N(h)(\mathbf{x_0}) \triangleq B(h)(\mathbf{x_0}) = \int_{\mathbb{T}^2} h \, d\mu,
\]
where $\mu$ is an invariant probability measure for the trajectory's closure. Due to edge effects at the two ends of the finite orbit segment, the convergence of \eqref{2} may be very slow. In \cite{kachurovskii1996rate}, it is shown that even if the orbit lies on a smooth invariant torus with irrational rotation number, the convergence rate of \eqref{2} is at best as \( \mathcal{O}(N^{-1})\). By contrast, for chaotic orbits, the convergence is typically considered to be $\mathcal{O}(N^{-1/2})$, which is essentially implied by the central limit theorem \cite{das2017quantitative, levnajic2010ergodic}. Although the convergence rate of the Birkhoff average can distinguish chaotic from regular orbits, the practical difference is often not significant enough for efficient classification.

To accelerate the convergence, we employ the weighted Birkhoff average (WBA) method. Following \cite{das2016quasiperiodicity}, the WBA is defined as
\begin{equation}
	WB_N(h)(\mathbf{x_0}) = \frac{1}{S} \sum_{n=0}^{N-1} \Psi \left( \frac{n}{N} \right) h(\mathbf{f}^n(\mathbf{x_0})), 
	\label{3}
\end{equation}
with the normalization constant
\[
S = \sum_{n=0}^{N-1} \Psi \left( \frac{n}{N} \right).
\]
$\Psi$ is a \( C^\infty \) weight function 
\begin{equation}
	\Psi(s) \triangleq
	\begin{cases}
		e^{-s(1-s)^{-1}}, & s \in (0,1), \\
		0, & \mathrm{else}
	\end{cases}
\end{equation}
to downplay the influence of the endpoints in a finite time series. This exponential bump function vanishes with infinite smoothness at the endpoints, meaning, $\lim\limits_{s \to 0^+} \Psi^{(k)}(s) = \lim\limits_{s \to 1^-} \Psi^{(k)}(s) = 0$ for all $k \in \mathbb{N}$. Consequently, we can define $\Psi^{(k)}(0) = \Psi^{(k)}(1) = 0$ for all $k \in \mathbb{N}$. It is established that 
\[
\lim_{N \to \infty} WB_N(h)(\mathbf{x_0}) = B(h)(\mathbf{x_0}).
\]

The key advantage of WBA lies in its convergence rate for regular orbits. For chaotic orbits, the convergence remains slow, with $ |WB_N(h)(\mathbf{x_0}) - B(h)(\mathbf{x_0})| \sim O(N^{-1/2}) $, similar to the standard Birkhoff average \cite{das2017quantitative, levnajic2010ergodic}. However, for regular orbits, the convergence can be dramatically faster \cite{das2017quantitative}. In particular, if the orbit is conjugate to a rigid rotation with a Diophantine rotation vector \( \omega \) and function \( h \) is \( C^\infty \), then \eqref{3} converges faster than any power, i.e., for any $k\in \mathbb{N}$, exists a constant $C_k$ such that
\[
|WB_N(h)(\mathbf{x_0}) - B(h)(\mathbf{x_0})| < \frac{C_k}{N^k}.
\]
Furthermore, if $ h $ is analytic, then the convergence rate is of exponential type $ \mathcal{O} \left( \exp(-N^\zeta) \right) $ for some $ \zeta>0 $ \cite{tong2024exponential}.
\begin{figure}[htbp]
	\centering
	\begin{subfigure}[b]{0.3\linewidth}
		\centering
		\includegraphics[width=1.1\linewidth]{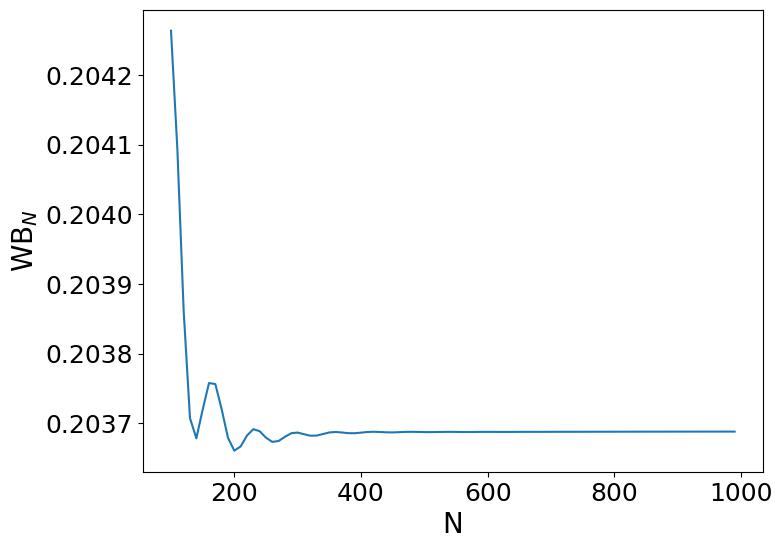}
		\caption{Non-resonant orbits}
		\label{fig5a}
	\end{subfigure}
	\hfill
	\begin{subfigure}[b]{0.3\linewidth}
		\centering
		\includegraphics[width=1.1\linewidth]{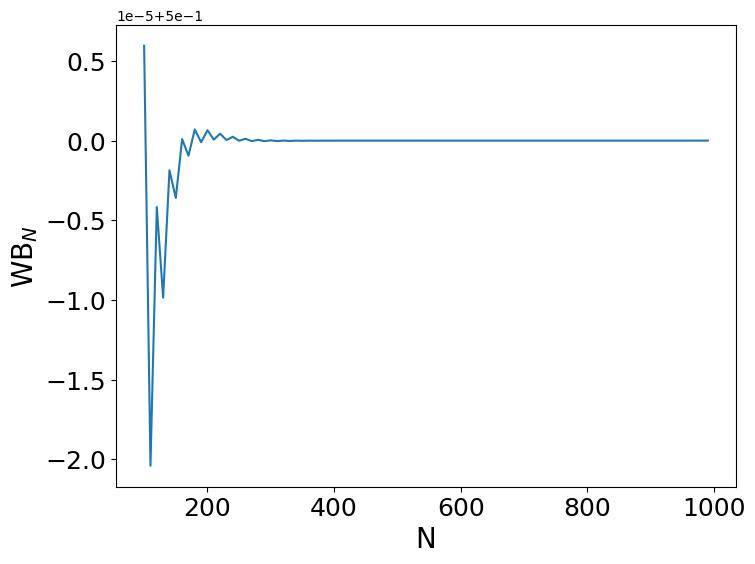}
		\caption{Resonant orbits}
		\label{fig5b}
	\end{subfigure}
	\hfill
	\begin{subfigure}[b]{0.3\linewidth}
		\centering
		\includegraphics[width=1.1\linewidth]{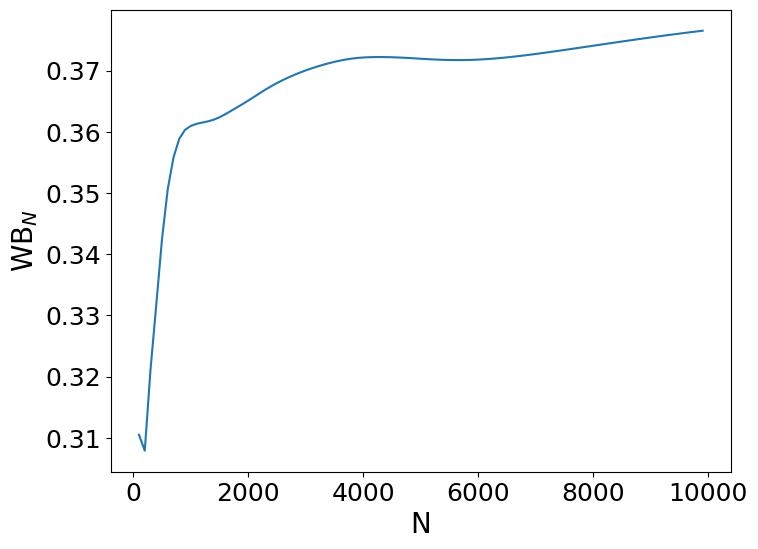}
		\caption{Chaotic orbits}
		\label{fig5c}
	\end{subfigure}
	\caption{Convergence behavior of the $WB_N$ series terms}
	\label{fig5}
\end{figure}

Figure \ref{fig5} illustrates the terms of the series $WB_N$ with $h(x,y) = y$  for three types of orbits in the GKR system. It is evident that for regular orbits (both non-resonant (a) and resonant orbits (b)), $WB_N$ converges rapidly, often within $1 \times 10^3$ iterations. In contrast, for chaotic orbit, no clear convergence is observed even after $2 \times 10^4$ iterations. This stark contrast demonstrates that analyzing the convergence of WBA provides an effective method for distinguishing chaotic from regular dynamics. However, since visually inspecting the convergence for every orbit is impractical, a quantitative criterion based on this method must be developed.

For a given orbit $\{\mathbf{f}^n(\mathbf{x_0}): n \in \mathbb{N}\}$, we compute the value of the $WB_N(h)$,  defined in Eq.\eqref{3} for two distinct segments of the orbit. The first segment comprises the initial $N$ iterates, $\left\{ \mathbf{x_0}, \mathbf{f}(\mathbf{x_0}), \mathbf{f}^2(\mathbf{x_0}), \dots,\mathbf{f}^{N-1}(\mathbf{x_0}) \right\}$. The second segment comprises the subsequent $N$ iterates, starting from the $N$-th point: $\left\{ \mathbf{f}^{N}(\mathbf{x_0}), \mathbf{f}^{N+1}(\mathbf{x_0}), \dots, \mathbf{f}^{2N-1}(\mathbf{x_0}) \right\}$. If the orbit is regular, $WB_N(h)$ converges rapidly. Consequently, the difference between its value computed over the first $N$ iterates and the next $N$ iterates should be negligible. In contrast, for a chaotic orbit, the discrepancy is typically significant. Based on this, we define a discriminating quantity:
\begin{equation}
	\Delta_N \triangleq WB_N(h)(\mathbf{x_0}) - WB_N(h)(\mathbf{f}^N(\mathbf{x_0})).
	\label{4}
\end{equation}

To classify the orbit, we quantify the magnitude of $\Delta_N$
by defining a measure of agreement between the two WBA estimates:
\begin{equation}
	\mathrm{dig}_N = -\log_{10} |\Delta_N|.
	\label{5}
\end{equation}
This metric, $\mathrm{dig}_N$, effectively measures the number of significant decimal digits to which the two estimates agree. For chaotic orbits, where $|\Delta_N| \sim N^{-1/2}$, $\mathrm{dig_N}$ remain relatively small. For regular orbits, both $WB_N(h)(\mathbf{x})$ and $WB_N(h(\mathbf{f}^N(\mathbf{x})))$ converge rapidly to the same limit, causing $\Delta_N$ to decay quickly to $0$; hence, $\mathrm{dig}_N$ becomes large and increases with $N$.

\begin{figure}[htbp]
	\centering
	\begin{subfigure}[b]{0.45\linewidth}
		\centering
		\includegraphics[width=1.1\linewidth]{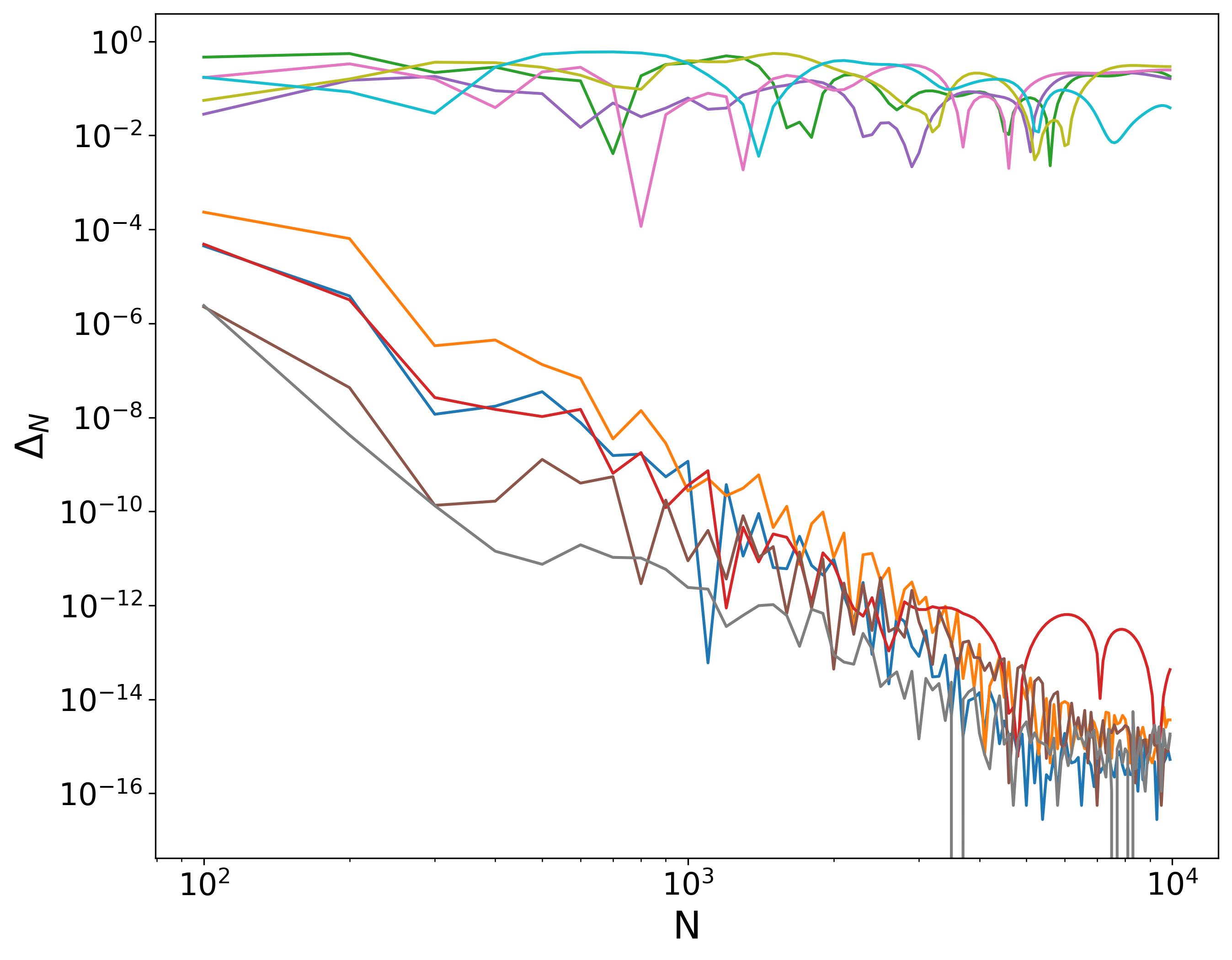}
		\label{fig6a}
	\end{subfigure}
	\hfill
	\begin{subfigure}[b]{0.45\linewidth}
		\centering
		\includegraphics[width=1.07\linewidth]{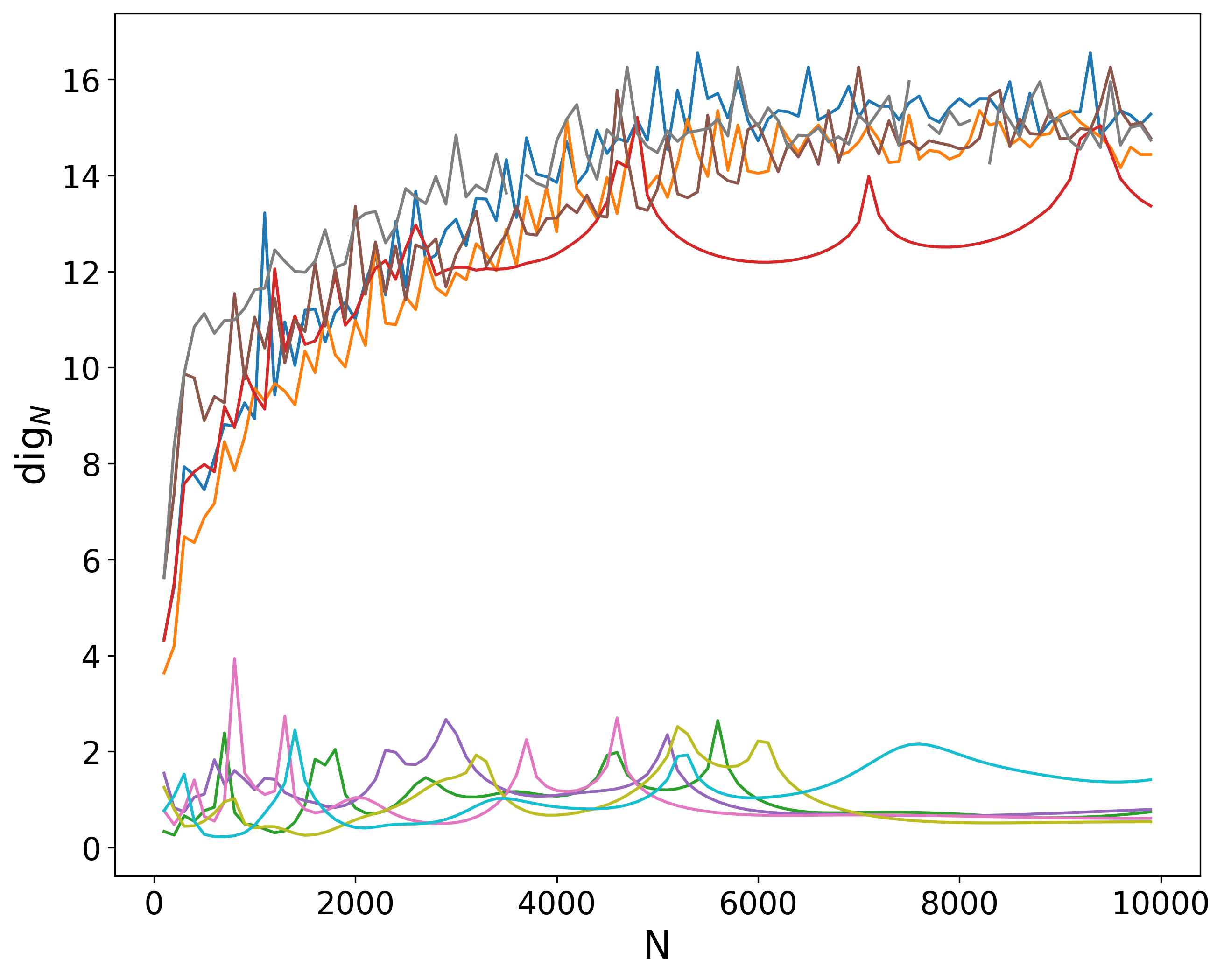}
		\label{fig6b}
	\end{subfigure}
	\caption{Left panel: $\Delta_N$ as a function of the number of iterations $N$ for 10 initial conditions. Right panel: $\mathrm{dig}_N$ as a function of $N$ for the corresponding orbits.}
	\label{fig6}
\end{figure}

Figure \ref{fig6} plots $\Delta_N$ against the number of iterations $N$ for ten orbits in the GKR system. For chaotic orbits, $\Delta_N$ remains largely unchanged as $N$ increase, however for resonant or non-resonant orbits, $\Delta_N$ decays rapidly. We can observe that there is a clear separation between chaotic and regular orbits for \( N \geq 10^4 \). Therefore, we choose \( N = 10^4 \) and use \( \mathrm{dig}_{10^4} \) as our classification metric.

\begin{figure}[h]
	\centering
	\begin{subfigure}[b]{0.45\linewidth}
		\centering
		\includegraphics[width=1.2\linewidth,height=3.4cm]{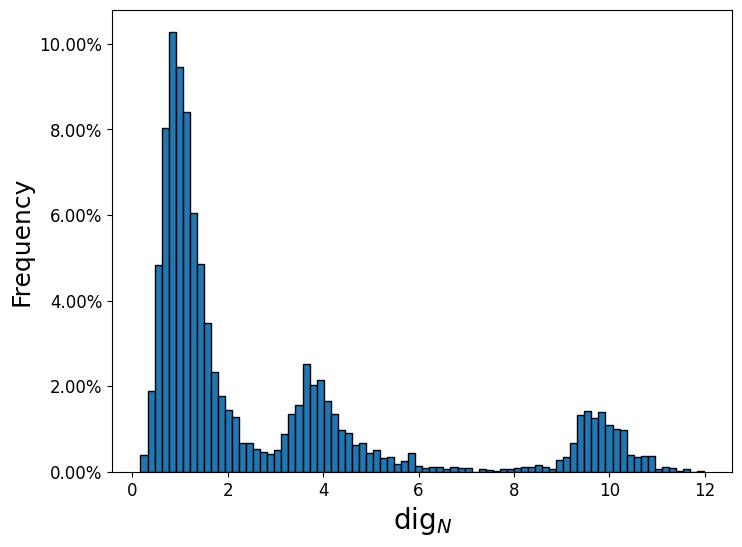}
		\label{fig7a}
	\end{subfigure}
	\hfill
	\begin{subfigure}[b]{0.45\linewidth}
		\centering
		\includegraphics[width=\linewidth]{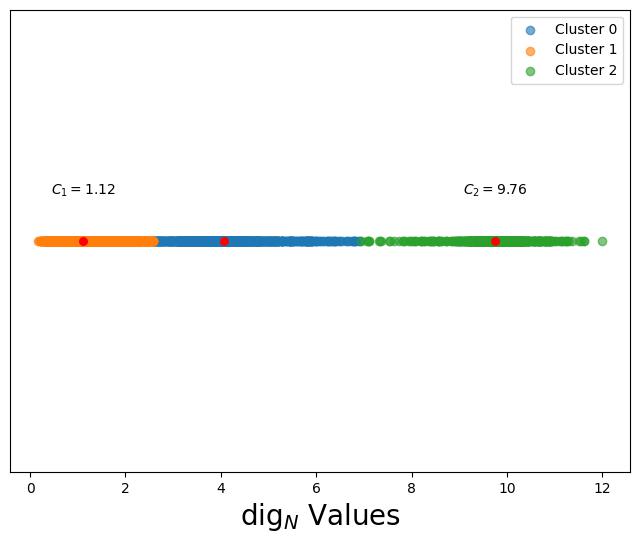}
		\label{fig7b}
	\end{subfigure}
	\caption{Left panel: Histogram of $\mathrm{dig}_{10^4}$ values for 5000 randomly selected initial conditions in the system Eq.\eqref{1} with $K=0.03$, $M=3$. Right panel: The corresponding clustering result of the $\mathrm{dig}_{10^4}$ data.}
	\label{fig7}
\end{figure}

Inspired by \cite{sander2020birkhoff}, we establish a threshold to distinguish orbits types. We performed a histogram analysis of \( \mathrm{dig}_{10^4} \) for $5000$ different initial conditions randomly selected from \( [0,1] \times [0,1] \), as shown in Figure \ref{fig7}. The distribution of \( \mathrm{dig}_{10^4} \) values reveals the concentrated intervals: a low-value region, a high-value region, and a transitional intermediate interval. We can consider the orbits in the low-value region to be strongly chaotic, and those in the high-value region to be regular. In fact, the Lyapunov exponents for orbits in the low-value region are significantly higher than those in the high-value region. Notably, within the transition region, Lyapunov exponents vary continuously. Some orbits show values close to zero (indicating regular orbits), others show high values (indicating strongly chaotic orbits), while another subset displays intermediate values. This phenomenon suggests the possible existence of weakly chaotic orbits, which warrants further investigation. Using $k$-$means$ clustering on the 
\( \mathrm{dig}_{10^4} \) values, we set thresholds \( C_1 \) (left) and \( C_2 \) (right) at the centroids of the low-value and high-value clusters. Orbits with \( \mathrm{dig}_{10^4} < C_1 \) are classified as strongly chaotic, and those with \( \mathrm{dig}_{10^4} > C_2 \) are classified as regular. Orbits falling between $C_1$ and $C_2$ (the transitional region) require further analysis.

As an example, for $K = 0.03$ and $M = 3$, the $\mathrm{dig}_{10^4}$ distribution (Figure \ref{fig7}) has the low-value and high-value centroids at $C_1=1.12$, $C_2=9.76$, respectively. We thus classify orbits with $\mathrm{dig}_{10^4} < 1.12$ as strongly chaotic (Figure \ref{fig13a}) and those with $\mathrm{dig}_{10^4} > 9.76$ as regular (Figure \ref{fig13c}) . Orbits with values between these thresholds are labeled as unclassified (Figure \ref{fig13b}). This procedure can be generalized to any parameters $K$ and $M$ using Algorithm~\ref{alg1} to systematically identify strongly chaotic and regular orbits.

\begin{algorithm}
	\caption{Stage \text{I}: Orbit Classification by weighted Birkhoff averaging}
	\begin{algorithmic}[1]
		\State \textbf{Input:} Parameters \( K \),\( M\), and orbits $\{(x_n, y_n)\}_{n=0}^{2\times 10^4}$
		\State \textbf{Output:} Regular \text{I}, Strongly chaotic \text{I}, and Unclassified \text{I}
		\For{each orbit}
		\State Calculate the $\mathrm{dig}_{10^4}$ using \eqref{5} 
		\EndFor
		\State Determine the two thresholds \( C_1 \) and \( C_2 \) by $k$-$means$ clustering
		\For{each orbit}
		\If{$\mathrm{dig}_{10^4} < C_1$}
		\State Labeled as Strongly chaotic \text{I} \ElsIf{$\mathrm{dig}_{10^4} > C_2$} 
		\State Labeled as Regular \text{I}
		\ElsIf{$C_1 \leq \mathrm{dig}_{10^4} \leq C_2$} 
		\State Labeled as Unclassified \text{I}
		\EndIf
		\EndFor
	\end{algorithmic}
	\label{alg1}
\end{algorithm}

\subsection{Lyapunov exponent}
\label{sec2.2}

In dynamical systems, the Lyapunov exponent (LE) is a key method for measuring a system's sensitivity to initial conditions, i.e., the exponential rate of divergence (or convergence) between nearby orbits. The Lyapunov exponent of the map $\mathbf{f}$ at the point $\mathbf{x_0}$ along the direction $\mathbf{v_0}$ is defined as:
\begin{equation}
	\lim_{N \to \infty}\frac{\ln \| D\mathbf{f}^N(\mathbf{x_0}) \mathbf{v_0} \|}{N}.
	\label{10}
\end{equation}
In practice, for finite but sufficiently large $N$, we compute the finite-time Lyapunov exponent using the formula \cite{das2017quantitative}:
\begin{equation}
	\frac{\ln \| D\mathbf{f}^N(\mathbf{x_0}) \mathbf{v_0} \|}{N} = \frac{1}{N} \sum_{n=0}^{N-1} \ln \| \mathbf{v_n} \|, \quad \mathbf{v_n} = \frac{D\mathbf{f}(\mathbf{x_{n-1}}) \mathbf{v_{n-1}}}{\| \mathbf{v_{n-1}} \|}. \quad 
	\label{6}
\end{equation}
Here, \( D\mathbf{f}(\mathbf{x}) \) denotes the Jacobian matrix of the map $\mathbf{f}$ evaluated at \( \mathbf{x} \). The sequence $\left\{\mathbf{v_n}\right\}$ represents the time-evolved perturbation vector. The Lyapunov exponent serves as a fundamental tool for characterizing chaotic behavior, providing crucial insights into the system's long-term evolution and predictability.

Note that for the GKR system \eqref{1}, the Jacobian matrix
\[
D\mathbf{f}(\mathbf{x}) =
\begin{bmatrix}
	1 + K \sum\limits_{n=1}^{M} 2\pi n \cos(2\pi n x) & 1 \\
	K \sum\limits_{n=1}^{M} 2\pi n \cos(2\pi n x) & 1
\end{bmatrix}
\] 
has a determinant $\det D\mathbf{f}(\mathbf{x_n}) \equiv 1$, implying area preservation in phase space. In order to compute the Lyapunov exponents, we choose two linearly independent initial perturbation directions, typically \( \mathbf{v_0}^{(1)}=(1, 0) \),  \( \mathbf{v_0}^{(2)}=(0, 1) \). Then, using Eq.\eqref{6} we obtain two Lyapunov exponents, denoted as \( \lambda_1 \) and \( \lambda_2 \). According to \cite{barreira2002lyapunov}, the sum of the Lyapunov exponents satisfies
\[
\lambda_1 + \lambda_2 = \lim_{N \to \infty} \frac{1}{N} \ln \left| \det D\mathbf{f}^N(\mathbf{x}) \right| = 0,
\]
where the term \( \left| \det D\mathbf{f}^N(\mathbf{x}) \right| \) quantifies the cumulative expansion or contraction of an infinitesimal phase space volume over $N$ iterations. Since $\lambda_1$, $\lambda_2$ are equal in magnitude and opposite in sign, we only need to consider either $|\lambda_1|$ or $|\lambda_2|$ (denoted hereafter as $|\lambda|$). In general dynamical systems, regular orbits typically correspond to all Lyapunov exponents being non-positive, while chaotic orbits are characterized by at least one positive Lyapunov exponent. Weakly chaotic orbits may exhibit Lyapunov exponents approaching zero. However, since \( f \) is an area-preserving map, both regular and weakly chaotic orbits may yield Lyapunov exponents near zero. To enhance the discriminative power of the visualization, we apply a logarithmic transformation to the Lyapunov exponents, thereby amplifying hierarchical differences in their magnitudes.

\begin{figure}[H]
	\centering
	\begin{subfigure}[b]{0.3\linewidth}
		\centering
		\includegraphics[width=1.15\linewidth]{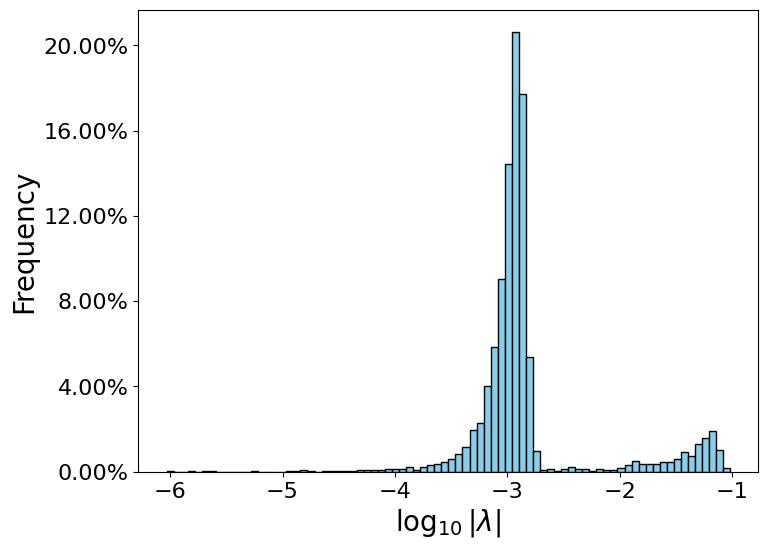}
		\caption{$K=0.01,M=3$}
		\label{fig8a}
	\end{subfigure}
	\hfill
	\begin{subfigure}[b]{0.3\linewidth}
		\centering
		\includegraphics[width=1.15\linewidth]{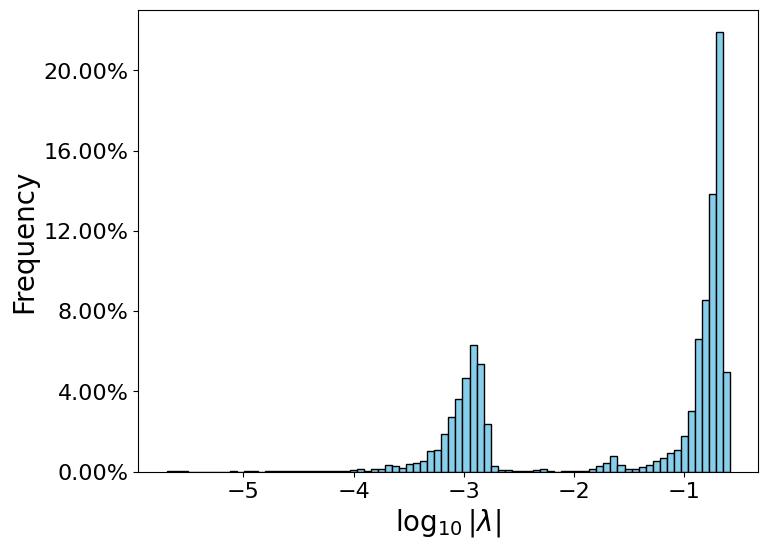}
		\caption{$K=0.03,M=3$}
		\label{fig8b}
	\end{subfigure}
	\hfill
	\begin{subfigure}[b]{0.3\linewidth}
		\centering
		\includegraphics[width=1.15\linewidth]{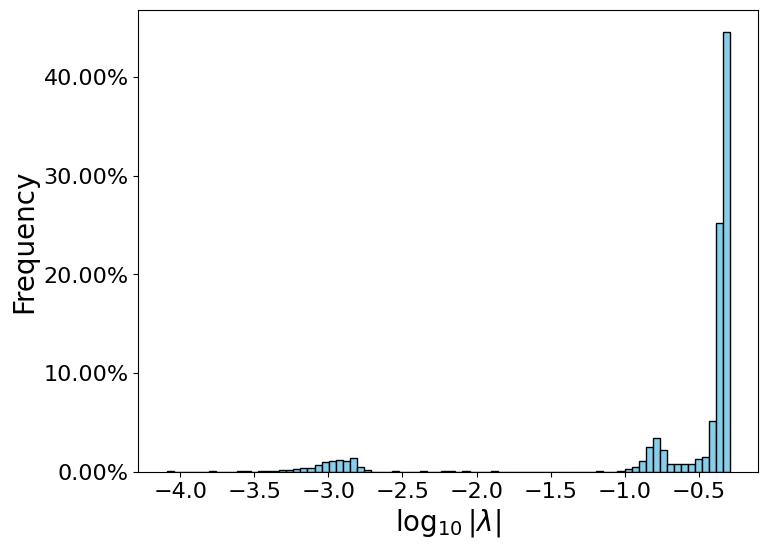}
		\caption{$K=0.1,M=3$}
		\label{fig8c}
	\end{subfigure}
	\vspace{1.0cm}
	\begin{subfigure}[b]{0.3\linewidth}
		\centering
		\includegraphics[width=1.15\linewidth]{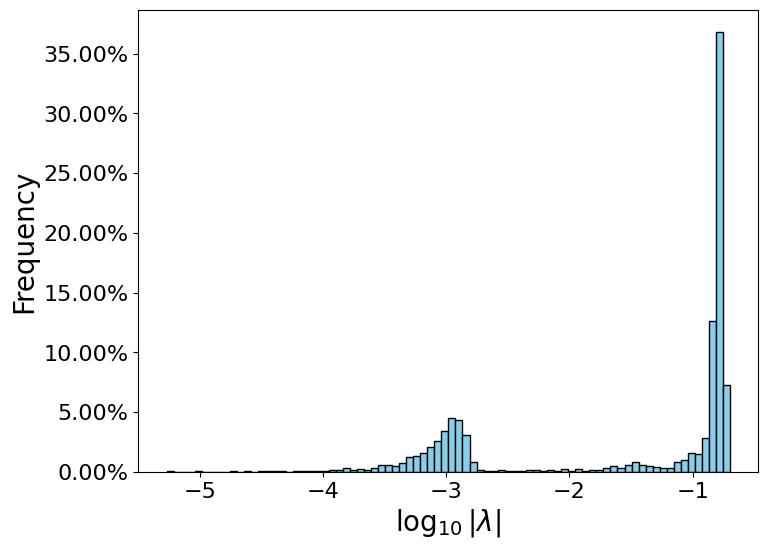}
		\caption{$K=0.01,M=5$}
		\label{fig8d}
	\end{subfigure}
	\hfill
	\begin{subfigure}[b]{0.3\linewidth}
		\centering
		\includegraphics[width=1.15\linewidth]{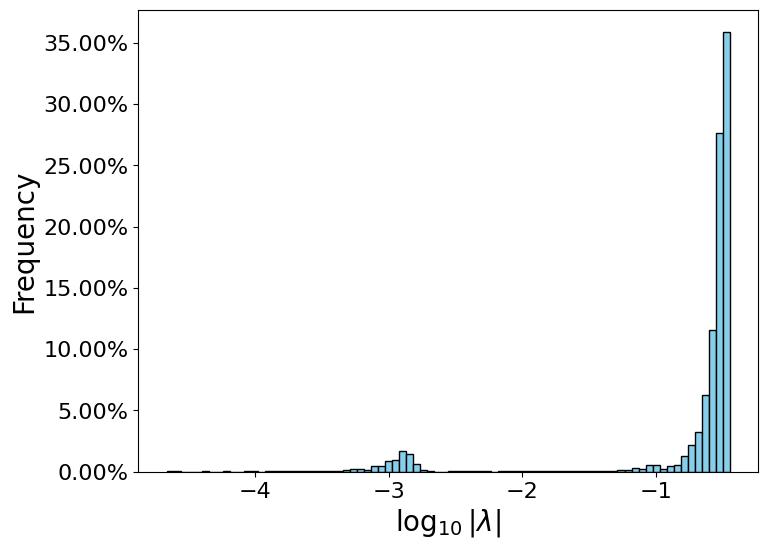}
		\caption{$K=0.03,M=5$}
		\label{fig8e}
	\end{subfigure}
	\hfill
	\begin{subfigure}[b]{0.3\linewidth}
		\centering
		\includegraphics[width=1.15\linewidth]{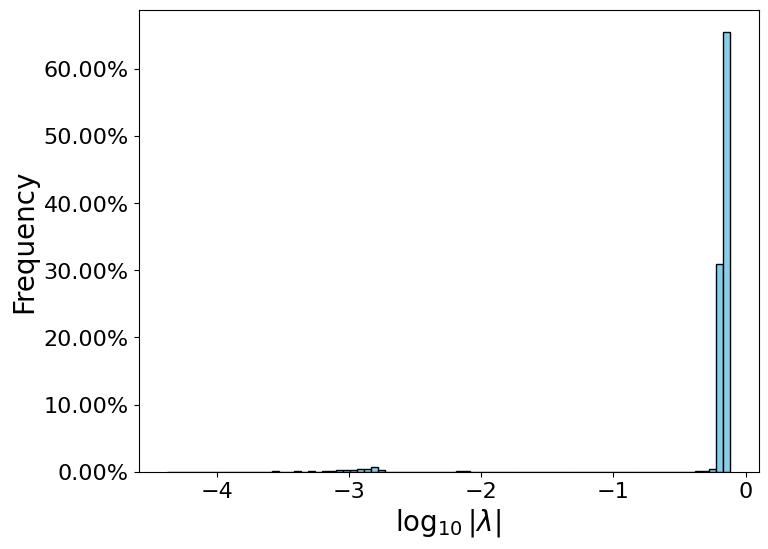}
		\caption{$K=0.1,M=5$}
		\label{fig8f}
	\end{subfigure}
	\caption{Histograms of \( \log_{10} |\lambda| \) for orbits of Eq.\eqref{1} with different $K$, $M$. The initial conditions are 5000 values randomly selected from the $[0,1] \times [0,1]$.}
	\label{fig8}
\end{figure}

We compute \(\log_{10} |\lambda| \) for orbits originating from $5000$ random initial conditions in \([0, 1] \times [0, 1]\), after $10^4$ iterations. Figure \ref{fig8} shows the corresponding histograms for different parameter sets: the top row for $M = 3$ with varying $K$, and the bottom row for $M = 5$. For \( K = 0.01 \) and \( M = 3 \), the distribution of \( \log_{10} |\lambda| \)  is primarily concentrated around \( -3 \), accompanied by an incipient secondary mound in the $\left[-2,-1\right]$ interval. As \( K \) increases, the peak 
\noindent of \( \log_{10} |\lambda| \) around \( -3 \) gradually weakens, while the feature around $-1$ grows significantly, eventually forming a dominant peak. At \( K = 0.03 \), a lower peak centered near \( -3 \) coexists with a dominant peak near \( -0.8 \), with clear separation between the two modes.  As \( K \) increases to $K = 0.1$, the system becomes predominantly chaotic, with a single prominent peak. A similar evolutionary pattern is observed for $M=5$ as $K$ increases, demonstrating that the transition from regular to chaotic dynamics is consistent for different numbers of harmonics.

Based on the preceding analysis, the value of $\log_{10} |\lambda|$ exhibit a consistent pattern: those of regular orbits concentrate on the left side of the histogram, while values from strongly chaotic orbits cluster on the right. This clear separation in $\log_{10} |\lambda|$ between the two previously identified groups provides a basis for further classifying the orbits in the Unclassified I category. To leverage this, we developed Algorithm \ref{alg2} as an extension of Algorithm \ref{alg1}. The procedure is as follows. We first compute $\log_{10}|\lambda|$ for the orbits already classified as Regular I and Strongly Chaotic I. From these, we define two thresholds: 
\begin{itemize}
	\item $\mathrm{LE}_{\mathrm{reg}}$: a value exceeds $\log_{10}|\lambda|$ for over 95\% of the Regular I orbits (i.e., $\log_{10}|\lambda| < \mathrm{LE}_{\mathrm{reg}}$).
	\item $\mathrm{LE}_{\mathrm{chaos}}$: a value that is exceeded by $\log_{10}|\lambda|$ for over 95\% of the Strongly chaotic I orbits (i.e., $\log_{10}|\lambda| > \mathrm{LE}_{\mathrm{chaos}}$).
\end{itemize}
These thresholds are then applied to the Unclassified I orbits. Those satisfying $\log_{10}|\lambda| < \mathrm{LE}_{\mathrm{reg}}$ are labeled as Regular II, as their Lyapunov exponents are consistent with Regular I group. Orbits satisfying $\log_{10}|\lambda| > \mathrm{LE}_{\mathrm{chaos}}$ are labeled Strongly chaotic II, indicating dynamics similar to Strongly chaotic I. 

\begin{algorithm}[htbp]
	\caption{Stage \text{II}: Identify orbits in Unclassified I by Lyapunov exponents}
	\begin{algorithmic}[1]
		\State \textbf{Input:} Parameters \( K \),\( M \), and orbits classified by Stage I 
		\State \textbf{Output:} Regular \text{II}, Strongly chaotic \text{II}, and Unclassified \text{II}
		\For{Regular \text{I} and Strongly chaotic \text{I}}
		\State Calculate the $\log_{10}|\lambda|$
		\EndFor
		\State Determine the thresholds $\mathrm{LE}_{\mathrm{reg}}$ and $\mathrm{LE}_{\mathrm{chaos}}$ such that:
		\Statex \hspace{0.5cm} For orbits in Regular I: $\Pr(\mathrm{LE}_{\mathrm{reg}} >\log_{10}|\lambda|) \geq 0.95$
		\Statex \hspace{0.5cm} For orbits in Strongly chaotic I: $\Pr(\mathrm{LE}_{\mathrm{chaos}}<\log_{10}|\lambda|) \geq 0.95$
		
		\For{Orbits in Unclassified \text{I}}
		\State Calculate the $\log_{10}|\lambda|$
		\If{$\log_{10}|\lambda| < \mathrm{LE}_{\mathrm{reg}}$}
		\State Labeled as Regular \text{II}
		\ElsIf{$\log_{10}|\lambda| > \mathrm{LE}_{\mathrm{chaos}}$}
		\State Labeled as Strongly chaotic \text{II}
		\ElsIf{$\mathrm{LE}_{\mathrm{reg}}\leq \log_{10}|\lambda| \leq \mathrm{LE}_{\mathrm{chaos}}$}
		\State Labeled as Unclassified \text{II}
		\EndIf
		\EndFor
	\end{algorithmic}
	\label{alg2}
\end{algorithm}		

\begin{figure}[htbp]
	\centering
	\begin{subfigure}[b]{0.45\linewidth}
		\centering
		\includegraphics[width=\linewidth]{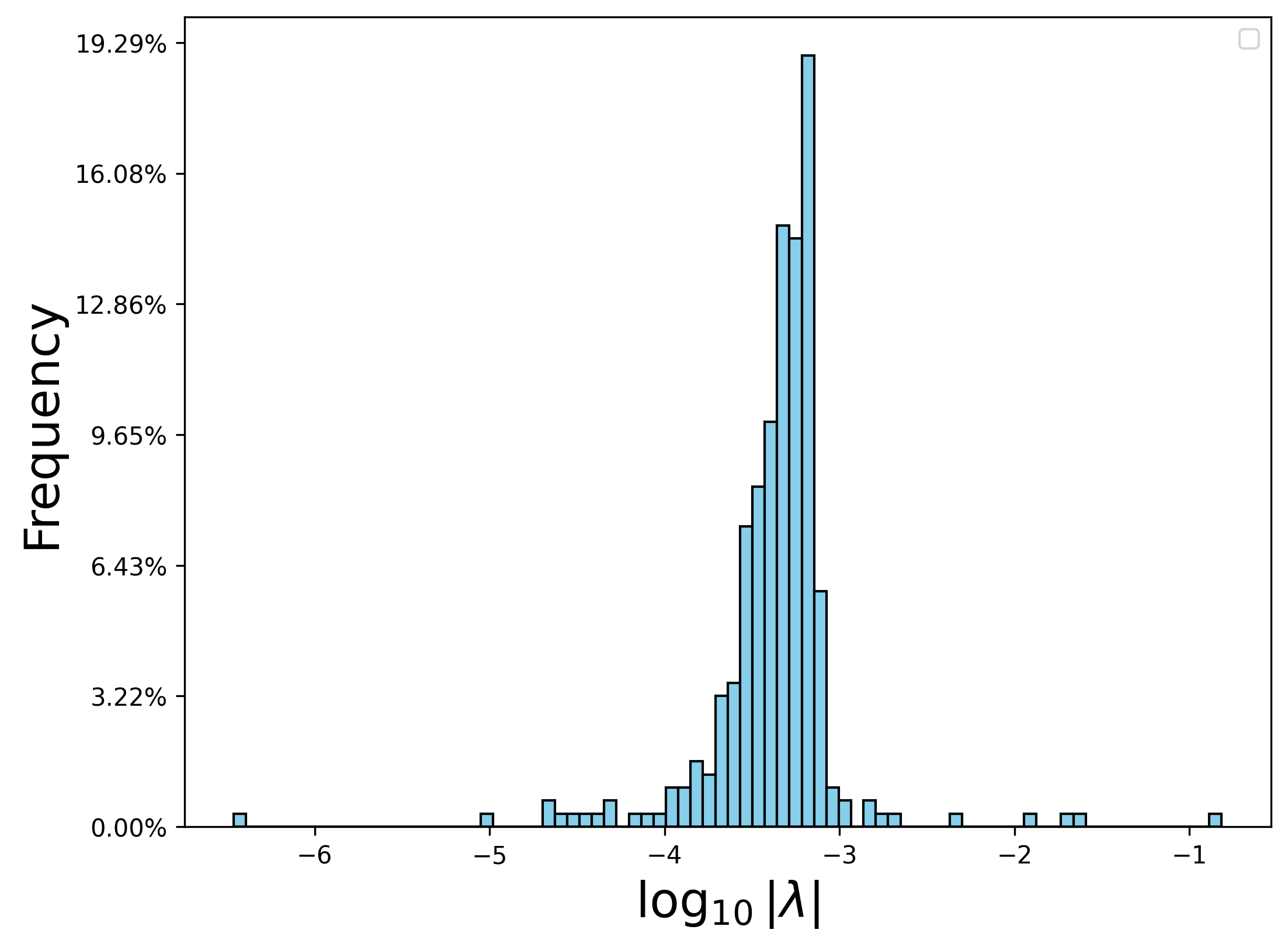}
	\end{subfigure}
	\hfill
	\begin{subfigure}[b]{0.45\linewidth}
		\centering
		\includegraphics[width=\linewidth]{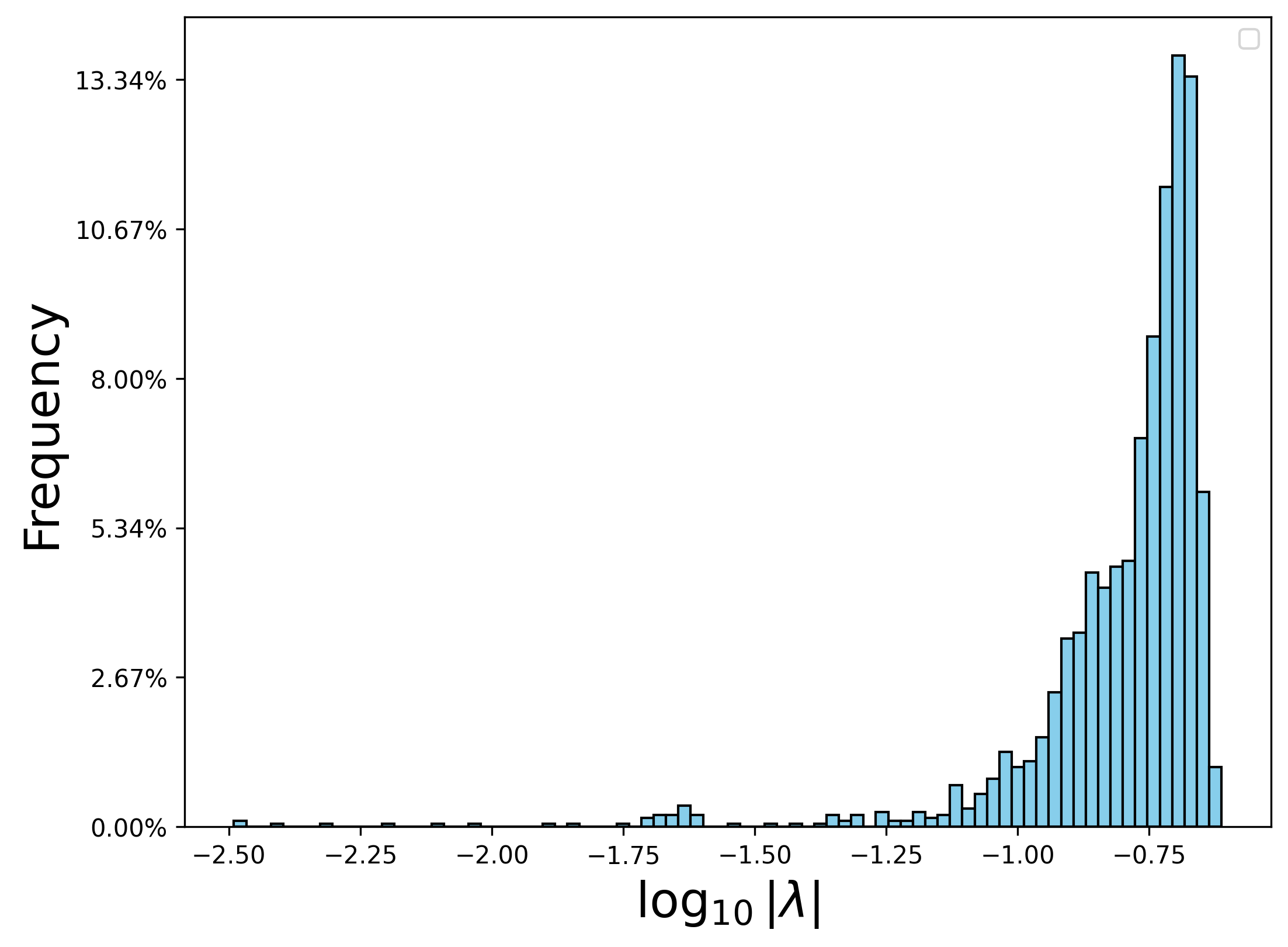}
	\end{subfigure}
	\caption{Histograms of \( \log_{10} |\lambda| \) for orbits with \( \mathrm{dig}_{10^4} > 9.76 \) (left), and \( \mathrm{dig}_{10^4} < 1.12 \) (right).}
	\label{fig9}
\end{figure}

For the specific case with $K = 0.03$ and $M = 3$, the orbits pre-classified in Algorithm \ref{alg1} exhibit distinct Lyapunov exponent distributions. As shown in Figure \ref{fig9}, regular orbits ($\mathrm{dig}_{10^4} > 9.76$) typically have $\log_{10}|\lambda| < -3.1$, with over 95\% of orbits falling below this value. In contrast, strongly chaotic orbits ($\mathrm{dig}_{10^4} < 1.12$) are characterized by $\log_{10}|\lambda| > -1.1$, a threshold exceeded by more than 95\% of orbits in this category. Therefore, we conclude that orbits in the ambiguous range $1.12 \leq \mathrm{dig}_{10^4} \leq 9.76$ that satisfy $\log_{10}|\lambda| < -3.1$ are classified as regular, while those with $\log_{10}|\lambda| > -1.1$ are classified as strongly chaotic. As shown in Figure \ref{fig13}, we further classify some orbits from the Unclassified I into strongly chaotic and regular orbits, as shown in Figures \ref{fig13d} and \ref{fig13f}, while a portion remains unclassified, as shown in Figure \ref{fig13e}.

\subsection{Correlation dimension}
\label{sec2.2.2}

The Correlation dimension (CD) is a fundamental measure in nonlinear time series analysis that quantifies the geometric complexity of a point set in a metric space by examining the statistics of pairwise distances. It serves as a powerful tool for characterizing the underlying structure of dynamical systems, particularly for assessing the effective degrees of freedom in complex systems and for identifying the fractal structure of strange attractors. The Grassberger-Procaccia (GP) algorithm is the most widely used method for computing the correlation dimension, especially for determining the fractal dimensions of strange attractors from observed time series data.

According to \cite{grassberger1983measuring}\cite{ding1993estimating}, for Eq.\eqref{1}, given an initial point $\mathbf{x_0}$, iterating \( N \) times yields an orbit from which the correlation sum \( C(N, r) \) can be computed as:
\begin{equation}
	C(N, r) = \frac{2}{N(N - 1)} \sum_{1 \leq i < j \leq N} H(r - \| \mathbf{x_i} - \mathbf{x_j} \|),
	\label{7}
\end{equation}
where \( H(\cdot) \) is the Heaviside step function and \( \|\mathbf{x_i} - \mathbf{x_j}\| \) denotes the Euclidean distance between embedded vectors. As \( r \to 0 \), \( C(N, r) \) decreases monotonically to zero. If \( C(N, r) \) follows a power law \( C(N, r) \sim r^{D} \) in the limits $N \to +\infty, r \to 0$, then \( D\) is referred to as the correlation dimension estimate of the orbit. This dimension is mathematically expressed as:
\begin{equation}
	D = \lim_{r \to 0} \lim_{N \to +\infty}\frac{\log C(N, r)}{\log r}.
	\label{8}
\end{equation}

\begin{figure}[H]
	\centering
	\begin{subfigure}[b]{0.45\linewidth}
		\centering
		\includegraphics[width=1.1\linewidth,height=3.2cm]{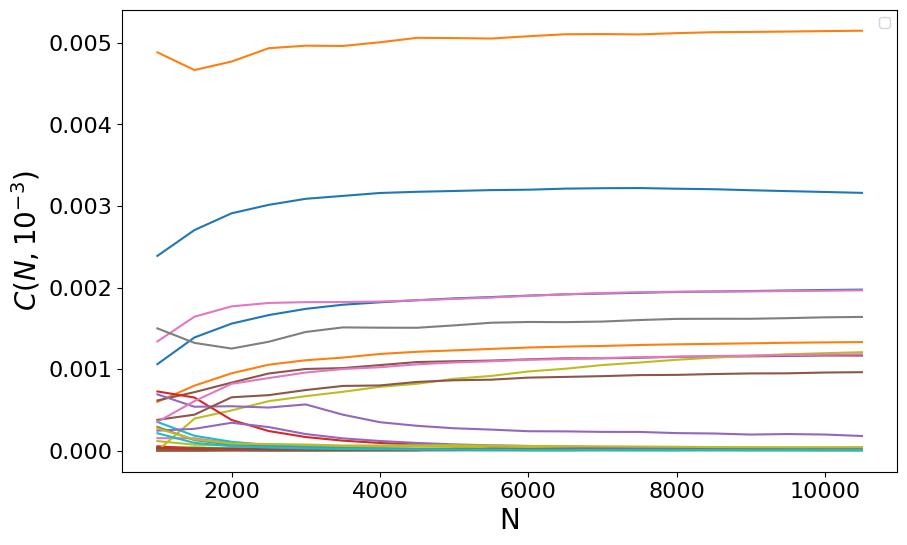}
		\label{fig10a}
	\end{subfigure}
	\hfill
	\begin{subfigure}[b]{0.45\linewidth}
		\centering
		\includegraphics[width=1.05\linewidth,height=3.2cm]{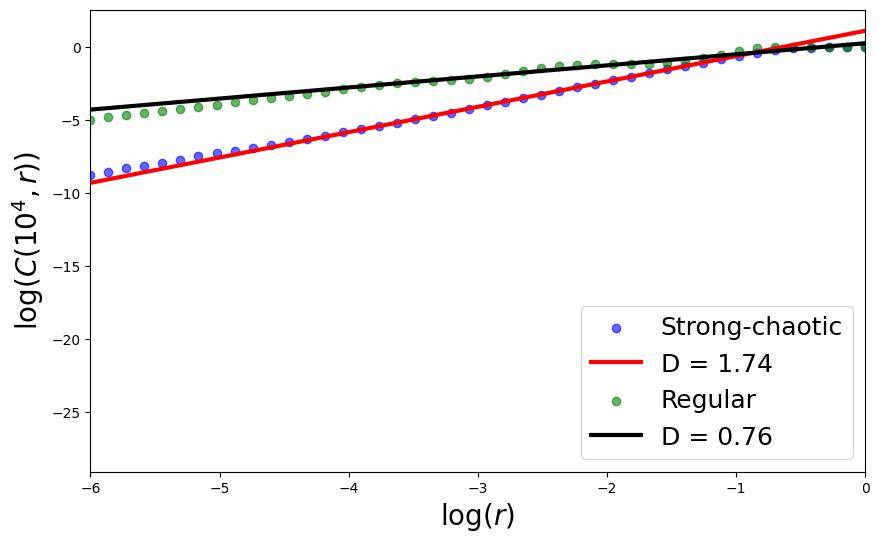}
		\label{fig10b}
	\end{subfigure}
	\caption{Left panel: $C(N, 10^{-3})$ as a function of the number of iterations N for 30 initial conditions. Right panel: Scatter plot of $\log C(10^4, r)$ against $\log r$ for orbits that were pre-classified into strongly chaotic and regular types. The black and red lines represent the fitted lines.}
	\label{fig10}
\end{figure}

The theoretical limits in Eq.\eqref{8} are not attainable in practice with finite data. Therefore, we adopt the following numerical method to estimate 
$D$. We generate orbits of length $N$ from 30 randomly selected initial conditions. To determine a suitable orbit length $N$, we examine the behavior of the correlation integral $C(N, r)$ as a function of  $N$ for a fixed radius $r$, with $r$ spanning the interval $[10^{-6}, 10^{0}]$. As shown in Figure \ref{fig10}, a similar trend was observed: $C(N, r)$ plateaus and becomes independent of $N$ for $N \geq 10^4$. Consequently, we set $N = 10^4$ for further computations. With $N$ fixed, we then evaluate $\log C(10^4, r)$ against $\log r$ over logarithmically spaced radii in the interval $[10^{-6}, 10^{0}]$. This range was chosen to cover the scaling region while avoiding saturation effects and noise, following the methodology of \cite{ding1993estimating, eckmann1985ergodic}. The correlation dimension $D$ was estimated from the slope of the linear region in the plot of $\log C(10^4, r)$ versus $\log r$. Figure \ref{fig10} presents scatter plots of orbits pre-classified as strongly chaotic and regular. The analysis reveals that regular orbit possess a lower correlation dimension than strongly chaotic orbit, aligning with the difference in their underlying dynamical complexity. 

\begin{figure}[H]
	\centering
	\begin{subfigure}[b]{0.3\linewidth}
		\centering
		\includegraphics[width=1.125\linewidth]{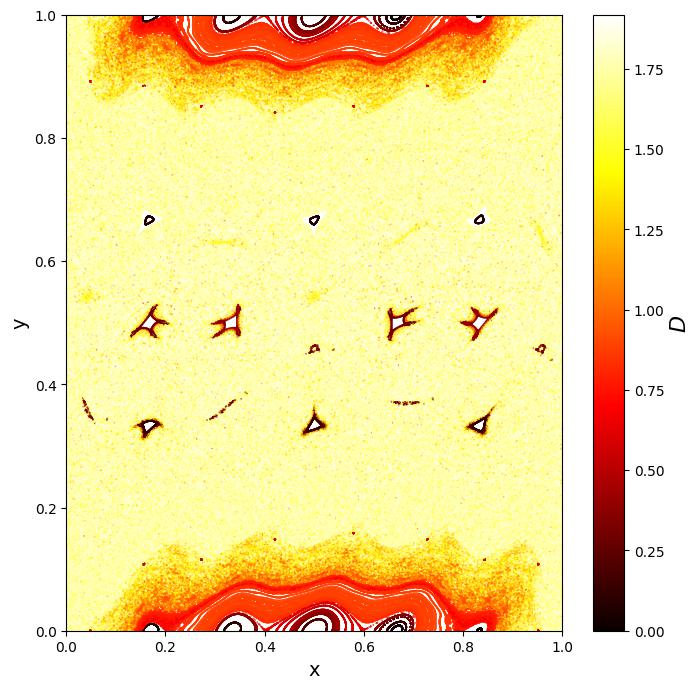}
		\caption{$K=0.03,M=5$}
		\label{fig11a}
	\end{subfigure}
	\hfill
	\begin{subfigure}[b]{0.3\linewidth}
		\centering
		\includegraphics[width=1.125\linewidth]{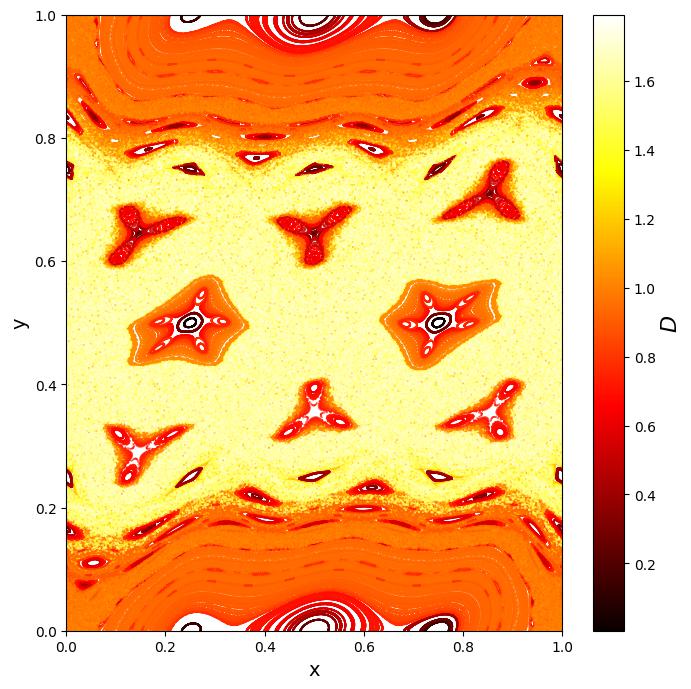}
		\caption{$K=0.03,M=3$}
		\label{fig11b}
	\end{subfigure}
	\hfill
	\begin{subfigure}[b]{0.3\linewidth}
		\centering
		\includegraphics[width=1.125\linewidth]{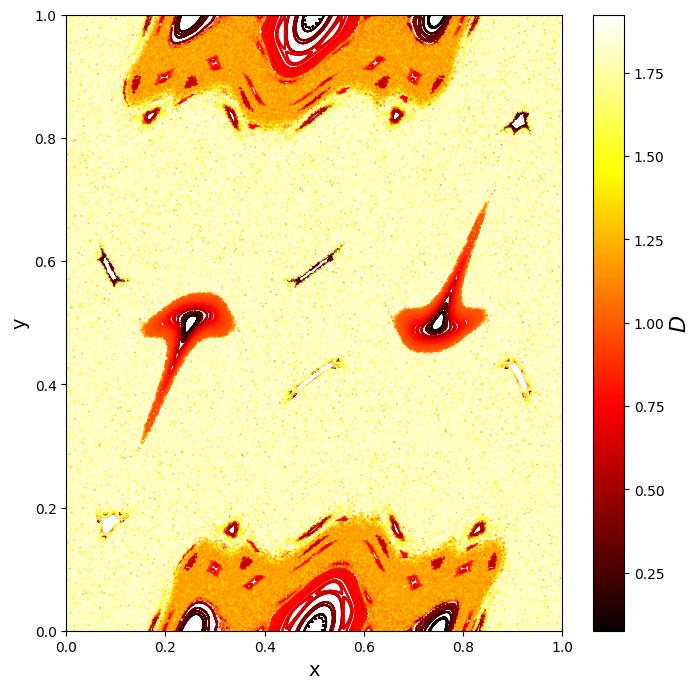}
		\caption{$K=0.1,M=3$}
		\label{fig11c}
	\end{subfigure}
	\caption{Correlation dimension heatmaps for Eq.\eqref{1}, computed from 400 orbits with initial conditions randomly sampled from the $[0,1] \times [0,1]$.}
	\label{fig11}
\end{figure}

To better visualize variations in the correlation dimension across a larger sample, we randomly select 400 initial conditions within $[0, 1] \times [0, 1]$ for different parameters $K$ and $M$. The corresponding distributions of $D$ are visualized using heatmaps.  As shown in Figure \ref{fig11}, these heatmaps illustrate the distribution of $D$ for specific $K$ and $M$, which consistently align with the distinct trajectory patterns previously identified in Figure \ref{fig1}. This finding demonstrates that the correlation dimension $D$, as a fundamental measure of set complexity, serves as an effective quantitative descriptor for distinguishing between different types of dynamical behaviors.

The preceding analysis reveals a consistent pattern in the correlation dimension $D$: it is significantly larger for strongly chaotic orbits than for regular ones. This pattern provides a quantitative basis for further classifying orbits in the Unclassified II category. To implement this, we developed Algorithm \ref{alg3}. The procedure begins by computing $D$ for orbits already categorized as Regular I $\cup$ II and Strongly Chaotic I $\cup$ II. Using these results, we define two thresholds:
\begin{itemize}
	\item $D_{\mathrm{reg}}$: a value exceeds $D$ for over 65\% of the Regular I $\cup$ II orbits (i.e., $D< D_{\mathrm{reg}}$).
	\item $D_{\mathrm{chaos}}$: a value that is exceeded by $D$ for over 65\% of the Strongly chaotic I $\cup$ II orbits (i.e., $D > D_{\mathrm{chaos}}$).
\end{itemize}
These thresholds are then applied to the Unclassified II orbits for classification:
\begin{itemize}
	\item Orbits with $D > D_{\mathrm{chaos}}$ are labeled as Strongly chaotic III, as their correlation dimensions align with the characteristic high values of the known strongly chaotic orbits. 
	\item Orbits with $D < D_{\mathrm{reg}}$ are labeled as Regular III, indicating dynamics similar to the known regular orbits. 
	\item Orbits with $D_{\mathrm{reg}} \leq D \leq D_{\mathrm{chaos}}$ are labeled as Weakly chaotic. 
	This intermediate classification is assigned to orbits whose behavior, as characterized by the synergistic integration of the three complementary methods---$\mathrm{dig}_{10^4}$, $\log_{10}|\lambda|$, and $D$---is neither unequivocally strongly chaotic nor regular.
\end{itemize}

\begin{algorithm}[H]
	\caption{Stage \text{III}: Identify orbits in Unclassified II by correlation dimensions}
	\begin{algorithmic}[1]
		\State \textbf{Input:} Parameters \( K \),\( M\), and orbits classified by Stage II
		\State \textbf{Output:} Regular, Strongly chaotic and Weakly chaotic orbits
		\For{\text{Regular I $\cup$ II} and \text{Strongly chaotic I $\cup$ II}}
		\State Calculate $D$ 
		\EndFor
		\State Determine the thresholds $D_{\mathrm{reg}}$ and $D_{\mathrm{chaos}}$ such that: 
		\Statex \hspace{0.5cm} For orbits in Regular I $\cup$ II: $\Pr(D_{\mathrm{reg}} > D) \geq 0.65$
		\Statex \hspace{0.5cm} For orbits in Strongly chaotic I $\cup$ II: $\Pr(D_{\mathrm{chaos}} < D) \geq 0.65$
		\For{Orbits in Unclassified \text{II}}
		\State Calculate $D$ 
		\If{$D < D_{\mathrm{reg}}$}
		\State Labeled as Regular \text{III} 
		\ElsIf{$D > D_{\mathrm{chaos}}$}
		\State Labeled as Strongly chaotic \text{III}
		\ElsIf{$D_{\mathrm{reg}} \leq D \leq D_{\mathrm{chaos}}$}
		\State Labeled as Weakly chaotic orbits
		\EndIf
		\EndFor
	\end{algorithmic}
	\label{alg3}
\end{algorithm}	

\begin{figure}[htbp]
	\centering
	\includegraphics[width=0.6\linewidth]{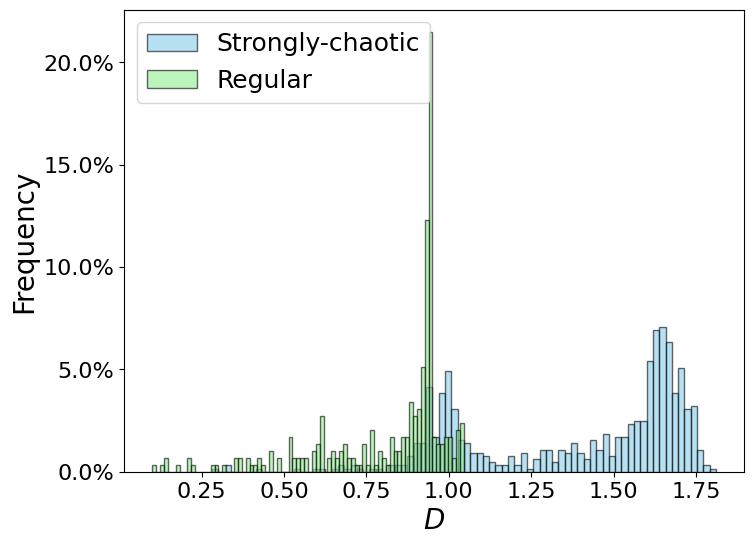}
	\caption{Histogram of the correlation dimension D. Skyblue and lightgreen bars represent the distributions for Strongly chaotic I $\cup$ II and Regular I $\cup$ II, respectively.
	}
	\label{fig12}
\end{figure}	

\begin{figure}[t]
	\centering
	\begin{subfigure}[b]{0.3\linewidth}
		\centering
		\includegraphics[width=1.1\linewidth]{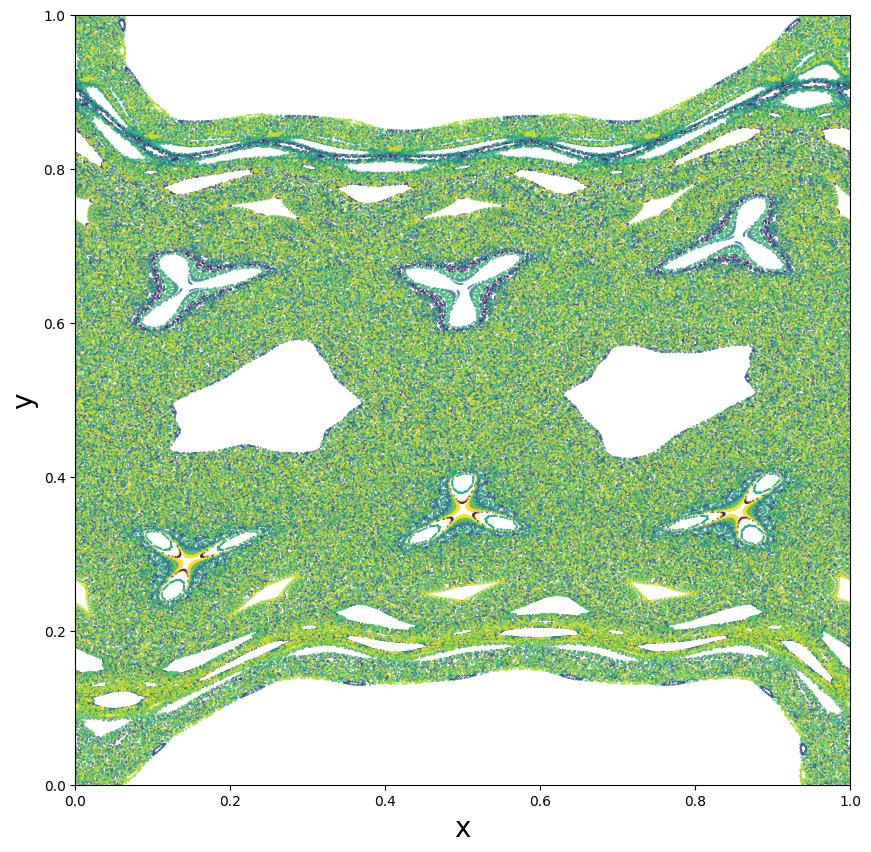}
		\caption{Strongly chaotic I}
		\label{fig13a}
	\end{subfigure}
	\hfill
	\begin{subfigure}[b]{0.3\linewidth}
		\centering
		\includegraphics[width=1.1\linewidth]{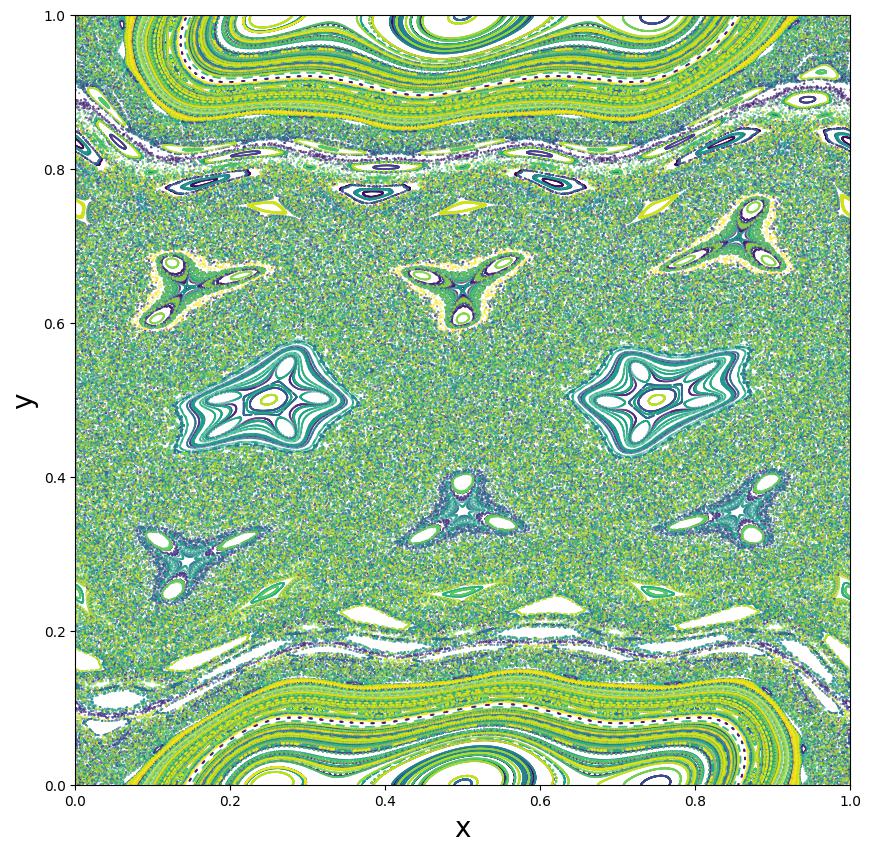}
		\caption{Unclassified I}
		\label{fig13b}
	\end{subfigure}
	\hfill
	\begin{subfigure}[b]{0.3\linewidth}
		\centering
		\includegraphics[width=1.1\linewidth]{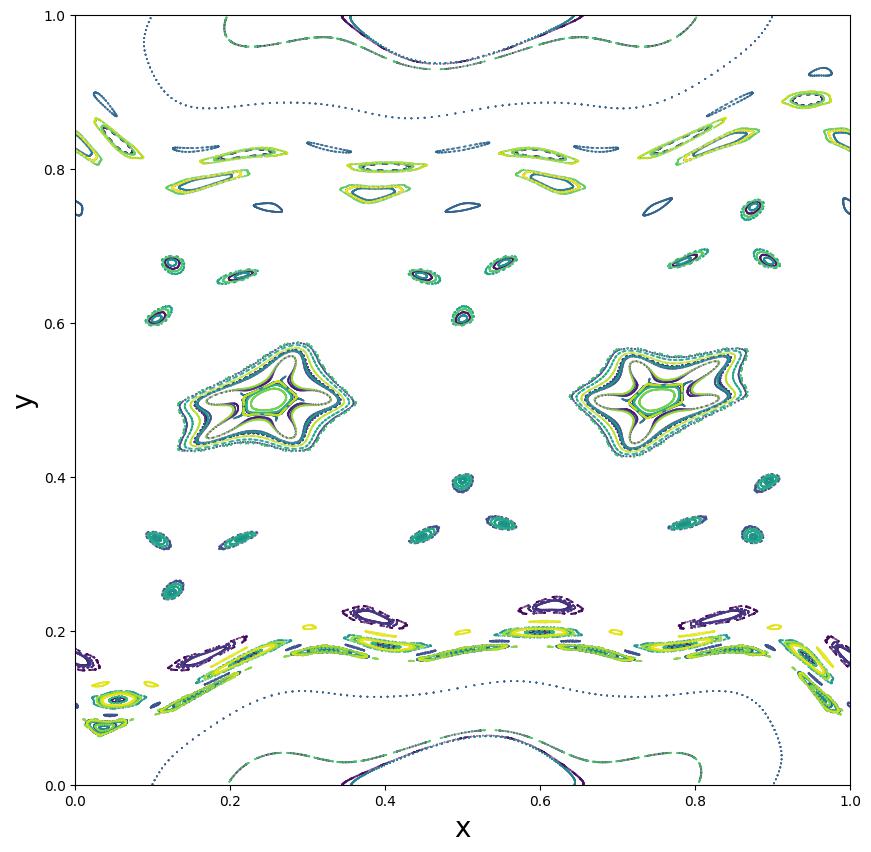}
		\caption{Regular I}
		\label{fig13c}
	\end{subfigure}
	
	\vspace{1em}
	
	\begin{subfigure}[b]{0.3\linewidth}
		\centering
		\includegraphics[width=1.1\linewidth]{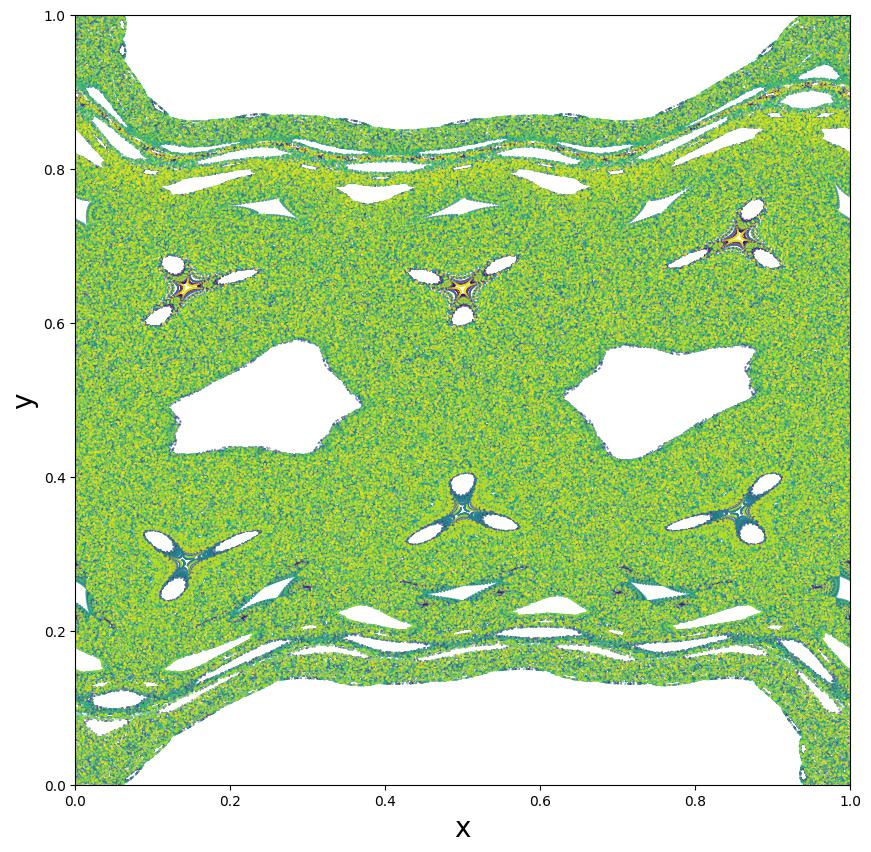}
		\caption{Strongly chaotic I $\cup$ II}
		\label{fig13d}
	\end{subfigure}
	\hfill
	\begin{subfigure}[b]{0.3\linewidth}
		\centering
		\includegraphics[width=1.1\linewidth]{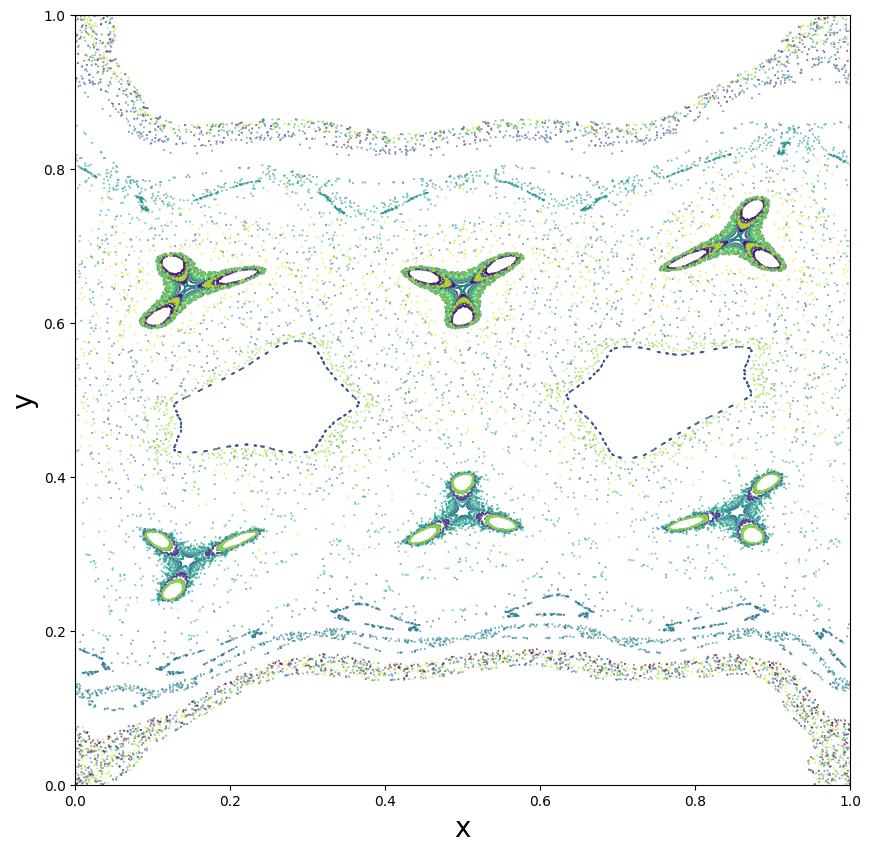}
		\caption{Unclassified II}
		\label{fig13e}
	\end{subfigure}
	\hfill
	\begin{subfigure}[b]{0.3\linewidth}
		\centering
		\includegraphics[width=1.1\linewidth]{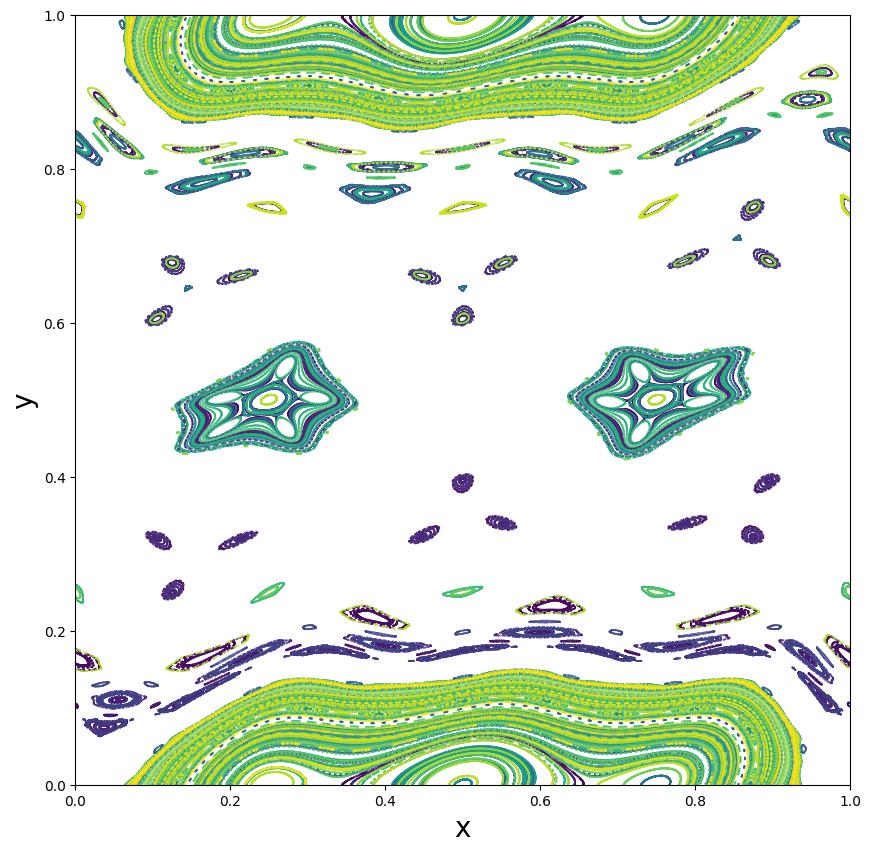}
		\caption{Regular I $\cup$ II}
		\label{fig13f}
	\end{subfigure}
	
	\vspace{1em}
	
	\begin{subfigure}[b]{0.3\linewidth}
		\centering
		\includegraphics[width=1.1\linewidth]{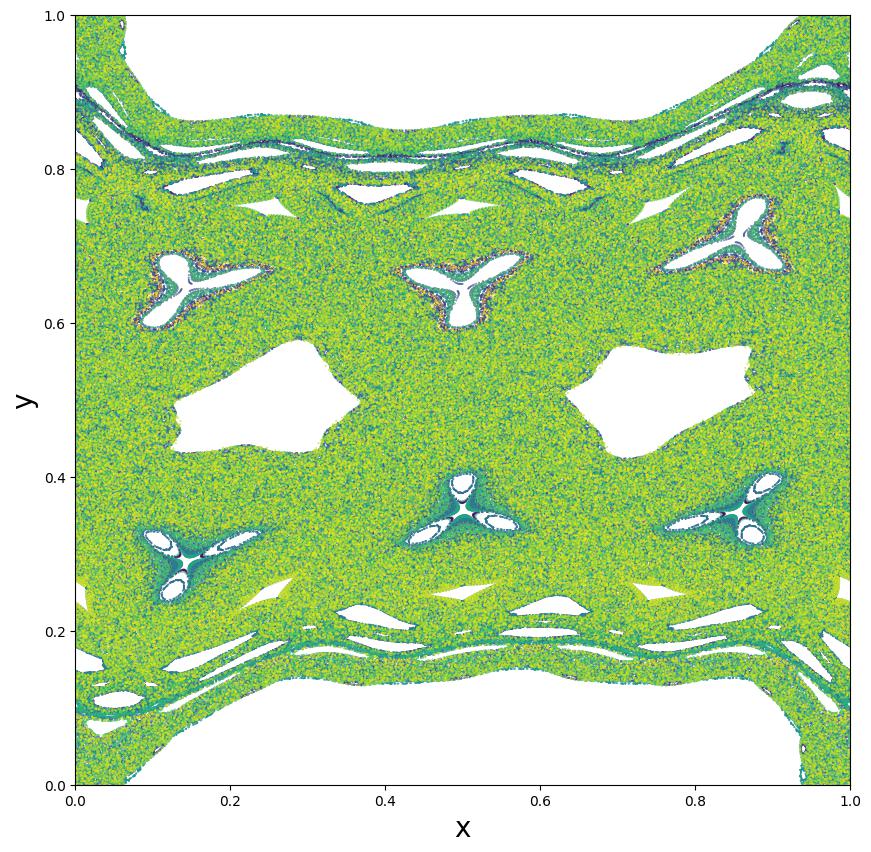}
		\caption{Strongly chaotic}
		\label{fig13g}
	\end{subfigure}
	\hfill
	\begin{subfigure}[b]{0.3\linewidth}
		\centering
		\includegraphics[width=1.1\linewidth]{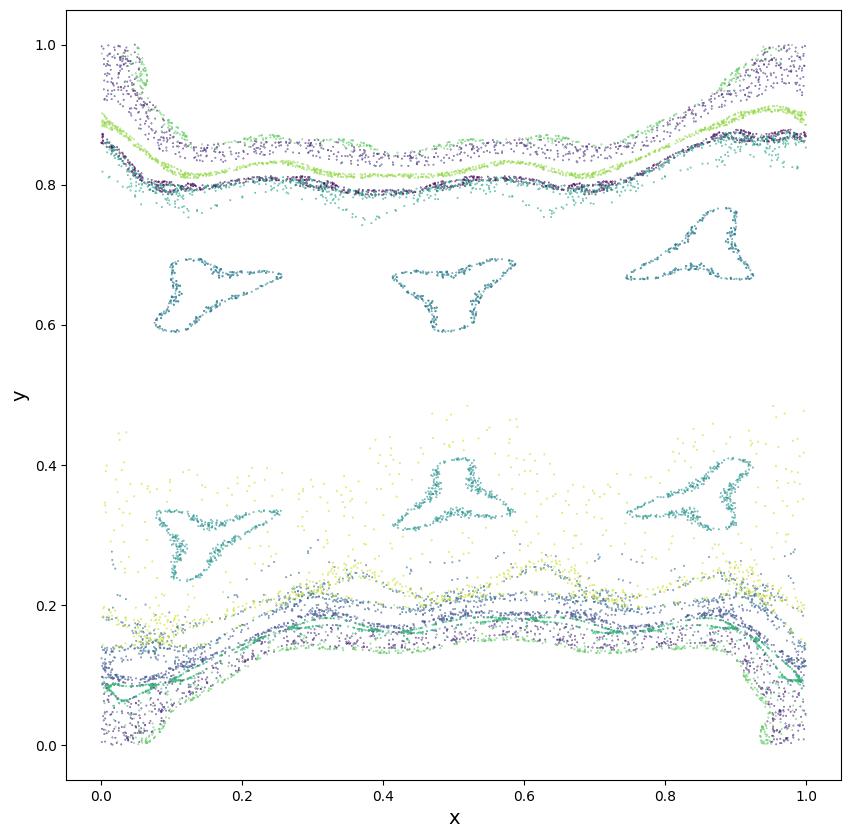}
		\caption{Weakly chaotic}
		\label{fig13h}
	\end{subfigure}
	\hfill
	\begin{subfigure}[b]{0.3\linewidth}
		\centering
		\includegraphics[width=1.1\linewidth]{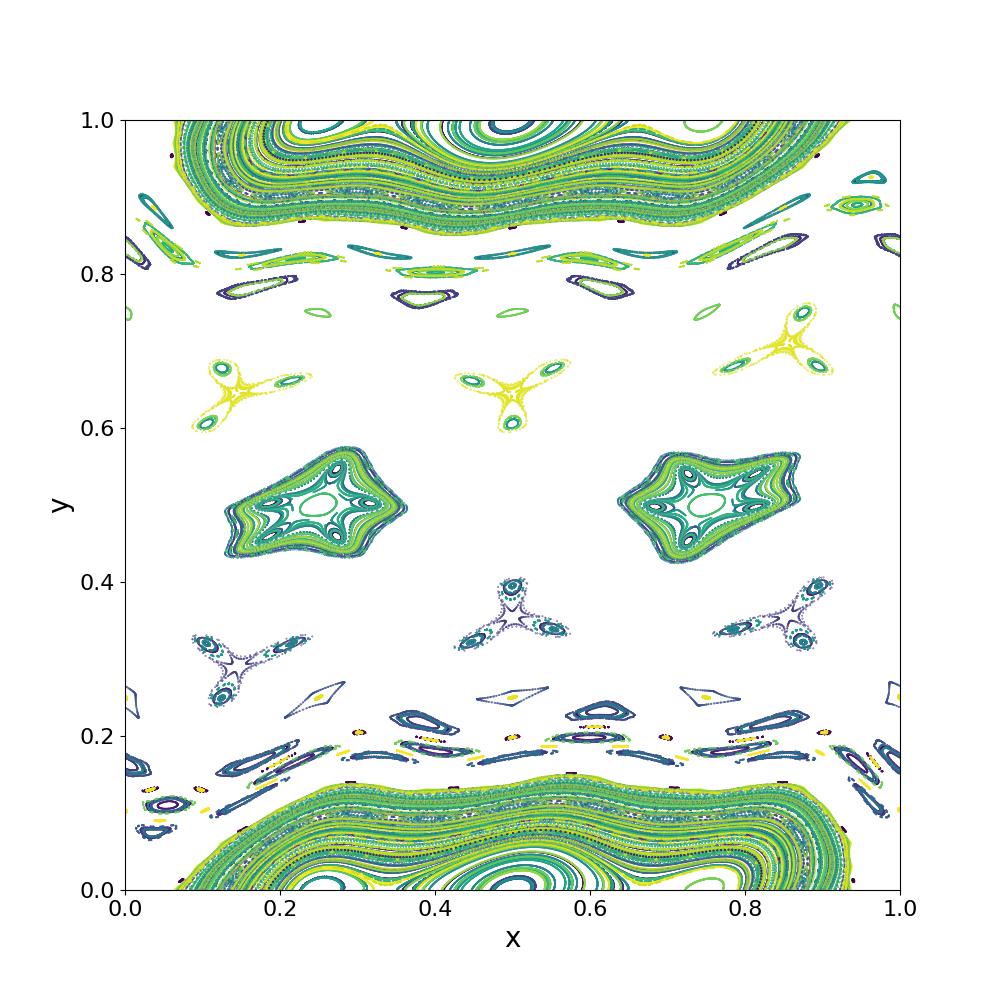}
		\caption{Regular}
		\label{fig13i}
	\end{subfigure}
	\caption{For the case with $K = 0.03$ and $M = 3$, the above algorithms progressively classify the orbits into: Strongly chaotic, Weakly chaotic, and Regular.}
	\label{fig13}
\end{figure}

For the specific case with \( K = 0.03 \) and \( M = 3 \), the orbits pre-classified in
Algorithm \ref{alg1} \ref{alg2} exhibit distinct correlation dimension distributions. As shown in Figure \ref{fig12}, more than 65\% of Strongly chaotic I $\cup$ II orbits have $D > 1.4$, while more than 65\% of Regular I $\cup$ II orbits have $D < 0.93$. Based on these thresholds, orbits in the Unclassified II set are classified as follows: those with $D > 1.4$ are classified as Strongly chaotic III, while those with $D < 0.93$ are classified as Regular III, and the remaining orbits are classified as Weakly chaotic. As visually summarized in Figures \ref{fig13g}, \ref{fig13h}, and \ref{fig13i}, this multi-stage hierarchical strategy successfully segregates all orbits into three distinct categories: Strongly Chaotic, Weakly Chaotic, and Regular.

\subsection{Identification of resonant invariant torus orbits}
\label{subsec2}

So far, we have categorized the orbits into three broad types---Strongly chaotic, Weakly chaotic, and Regular---using Algorithms \ref{alg1}, \ref{alg2}, and \ref{alg3}. We now focus on the regular orbits, which are further subdivided into resonant and non-resonant orbits using Algorithm \ref{alg4}.

Regular orbits consist of two topologically distinct types: resonant orbits and non-resonant orbits. A central quantity for characterizing these orbits is the rotation number $\omega$, defined for any initial condition on a regular orbit as:
\begin{equation}
	\omega = \lim_{N \to \infty} \frac{1}{N} \sum_{n=0}^{N-1} y_{n+1}.
	\label{9}
\end{equation}
This quantity quantifies the average angular rate of motion. If $\omega$ satisfies a strong irrationality condition, such as the Diophantine condition, the corresponding orbit is non-resonant, and its persistence under small perturbations is assured by the KAM theorem. Conversely, if $\omega$ is sufficiently well-approximated by rational numbers, the orbit is resonant, a structure typically destroyed by perturbations.

To numerically distinguish between these cases, we employ a method that assesses, with high probability, whether a floating-point computed $\omega$ approximates a rational or an irrational number \cite{sander2020birkhoff}. This method defines an indicator, $\text{dev}_\omega$, for each computed rotation number. If $\text{dev}_\omega > s$,  the rotation number is identified as approximating a rational number, indicating that the corresponding orbit is resonant; otherwise, it is considered an irrational approximation, meaning the orbit is non-resonant. In our implementation, the threshold is set to $s=0.67$(see \ref{Appendix A} for details). Applying Algorithm \ref{alg4} with parameters $K=0.03$ and $M=3$ successfully categorizes the orbits within the regular set into resonant and non-resonant types. Thus, we achieve a comprehensive four-fold classification of orbits, establishing a systematic framework that distinguishes strongly chaotic, weakly chaotic, resonant, and non-resonant orbits.

\begin{algorithm}
	\caption{Stage \text{IV}: Classify regular orbits by rotation number}
	\begin{algorithmic}[1]
		\State \textbf{Input:} Regular orbits
		\State \textbf{Output:} resonant orbits and non-resonant orbits 
		\For {orbits in Regular}
		\State Compute $\text{dev}_\omega$ 
		\If{$\text{dev}_\omega>0.67$}
		\State Labeled as resonant orbit
		\ElsIf{$\text{dev}_\omega \leq 0.67$}
		\State Labeled as non-resonant orbit
		\EndIf
		\EndFor	
	\end{algorithmic}
	\label{alg4}
\end{algorithm}

\begin{figure}[htbp]
	\centering
	\begin{subfigure}[b]{0.3\linewidth}
		\centering
		\includegraphics[width=1.2\linewidth]{9_7.jpg}
		\caption{Regular orbits}
		\label{fig14a}
	\end{subfigure}
	\hfill
	\begin{subfigure}[b]{0.3\linewidth}
		\centering
		\includegraphics[width=1.2\linewidth]{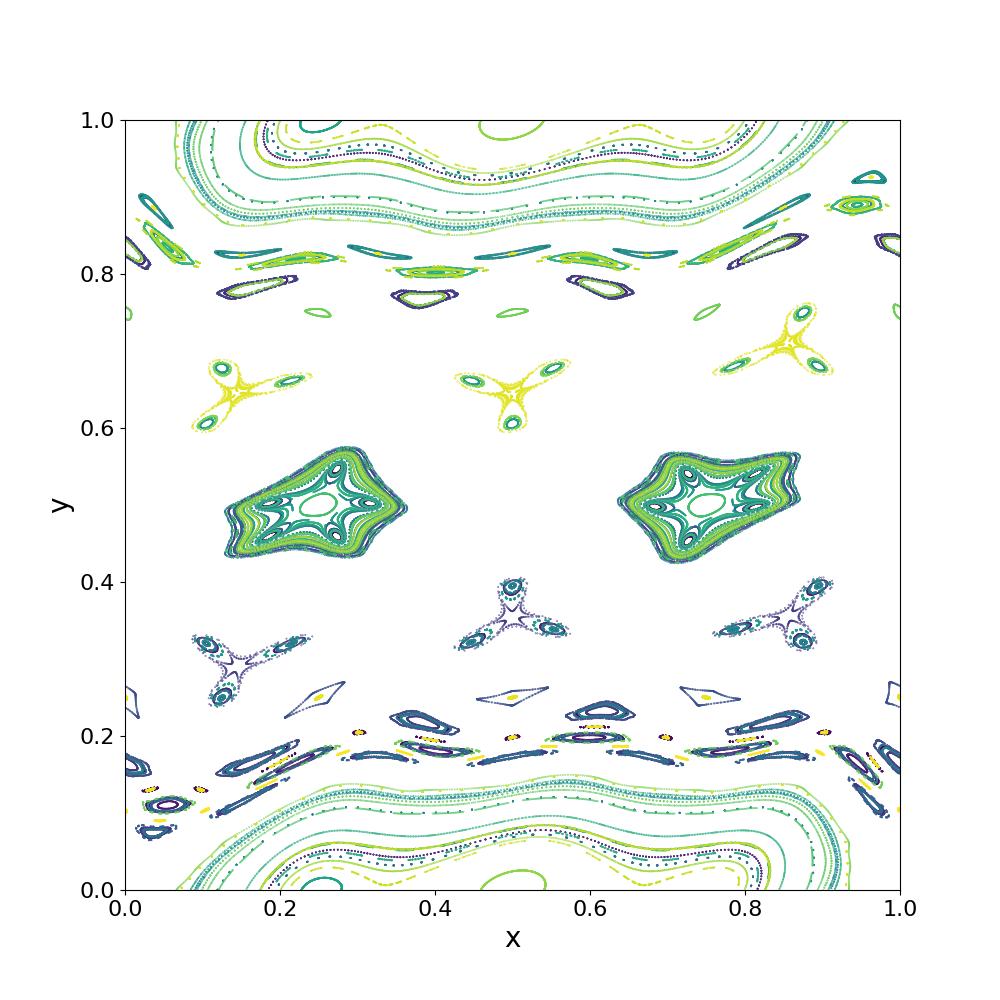}
		\caption{Resonant orbits}
		\label{fig14b}
	\end{subfigure}
	\hfill
	\begin{subfigure}[b]{0.3\linewidth}
		\centering
		\includegraphics[width=1.2\linewidth]{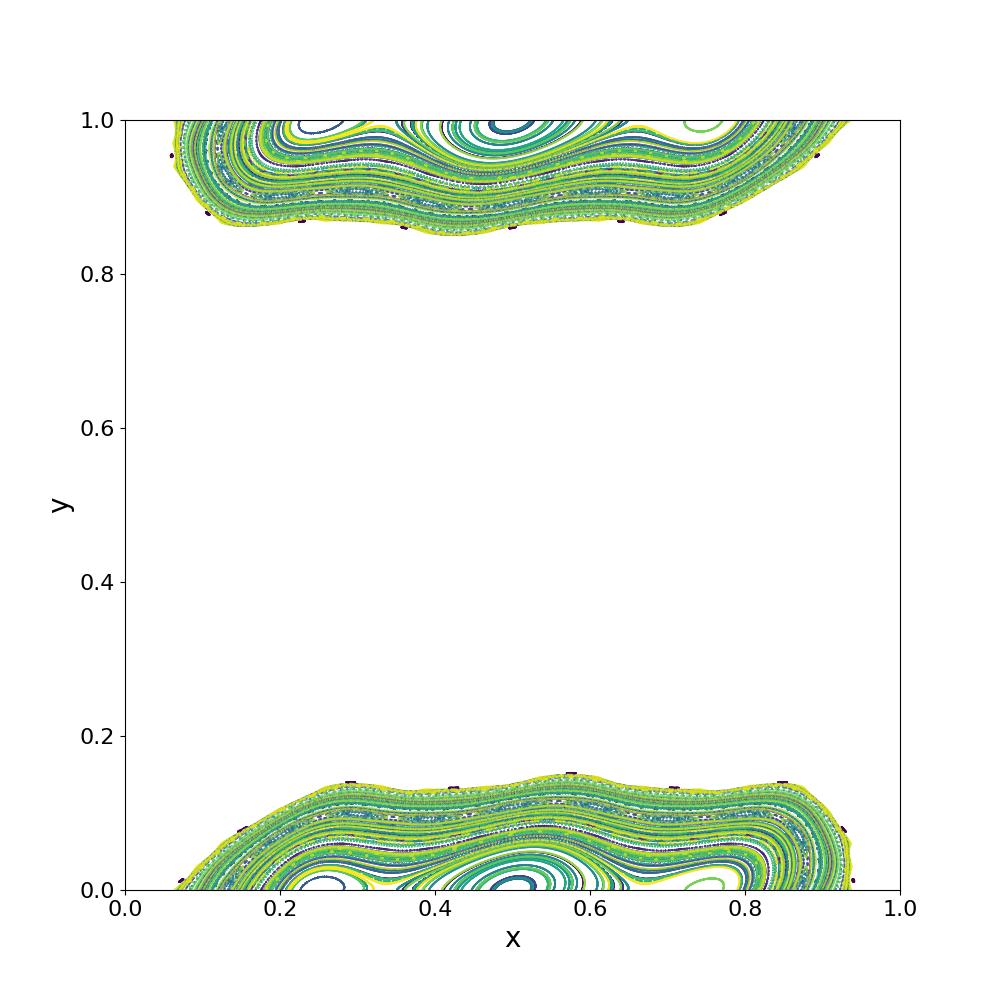}
		\caption{Non-resonant orbits}
		\label{fig14c}
	\end{subfigure}
	\caption{The regular orbits in the case of $K=0.03$, $M=3$ are further classified into resonant orbits and non-resonant orbits.}
	\label{fig14}
\end{figure}

\section{Deep Learning classification}
\label{sec3}

This section outlines the generation of high-quality labeled data for deep learning, specifies which deep learning models were selected, explains the training process used for orbit classification, and provides a comparative analysis of their classification performance. MLP is typically employed as the baseline model. InceptionTime CNN architecture has demonstrated strong classification performance in time-series classification tasks, as evidenced by previous related studies. Furthermore, we implement hybrid variants of the MLP and InceptionTime CNN that integrate Transformer architectures, resulting in the MLP-Transformer and InceptionTime CNN-Transformer models. The incorporation of Transformers is motivated by their self-attention mechanism, which excels at capturing complex long-range dependencies inherent in time-series data. Finally, we employ a 2D-CNN model that converts sequential orbital data into image-based representations. This approach enables the model to leverage the inherent spatial patterns in trajectory images for accurate orbit type identification. Among the models evaluated on the same dataset, the 2D-CNN demonstrates the most effective and stable classification performance.

\subsection{Details about the Data and Model}
\label{sec3.1}

\textbf{Datasets}: To generate high-quality labeled data for deep learning, we first evolve 1000 orbits from given initial conditions over long-term iterations. Each orbit is algorithmically assigned a definitive label: '0' for strongly chaotic orbits, '1' for resonant orbits, '2' for non-resonant orbits, and '3' for weakly chaotic orbits. These reliable labels are then paired with corresponding shorter 900-point orbital segments to construct the dataset, which is partitioned into training (64\%), validation (16\%), and test (20\%) sets. This strategy leverages long-term evolution for robust labeling and uses shorter segments as model inputs, effectively balancing label reliability with data diversity to establish a scalable and practical foundation for deep learning-based classification. 

\textbf{MLP and MLP-Transformer}: In classification tasks, MLP is typically employed as a benchmark model, trained on our specialized dataset derived from orbital trajectories. Each sample is constructed by concatenating two feature sets: a downsampled time series, created by selecting every 10th point from the original 900-length orbit and then flattened; and a vector of statistical descriptors, including the mean, minimum, maximum, variance, correlation coefficient, and average density. Our MLP architecture comprises an input layer, four fully-connected hidden layers (each containing 500 neurons), and an output layer with four neurons corresponding to the classification categories. Hidden layers utilize the ReLU activation function, batch normalization and dropout regularization ($p=0.1$).
The output layer employs a softmax function to generate class probabilities. The training process employs the AdamW optimizer with a weight decay coefficient of $10^{-5}$ and gradient clipping. The learning rate is scheduled using the OneCycle policy with a peak learning rate of 0.01. The loss function is cross-entropy, and the model is trained for 500 epochs. During training, we apply updates in batches of 64. The model achieving the highest validation accuracy is saved for final evaluation on the test set. Building upon this MLP baseline, we design an MLP-Transformer hybrid architecture that processes the input through dual parallel pathways.  In the Transformer pathway, the input is reshaped into a sequence and augmented with sinusoidal positional encoding, then processed by a 3-layer encoder with 8-headed self-attention, followed by adaptive average pooling, while the MLP pathway consists of two 256-neuron layers. Both pathways employ batch normalization, ReLU activation, and dropout. Their respective representations are concatenated and processed through three fusion layers (512, 256, and 4 neurons) utilizing identical regularization techniques. This hybrid model maintains the same optimization strategies and training protocols as the MLP baseline.

\textbf{InceptionTime CNN and InceptionTime CNN-Transformer}: Following the approach in \cite{celletti2022classification}, which applies the InceptionTime CNN architecture to classify regular and chaotic motions in the forced pendulum and the spin–orbit systems, our work employs a standard InceptionTime CNN model with six Inception modules. The network input is a two-channel time series representing the temporal evolution of the x- and y-coordinates over 900 iterations. Each Inception module begins with a bottleneck layer, followed by parallel branches: three convolutional layers with kernel sizes of 39, 19, and 9, and one max-pooling branch that is subsequently processed by $1\times1$ convolution. The outputs of these branches are concatenated, processed through batch normalization, and activated via the ReLU function. The final feature representation is subjected to global average pooling before being passed to a fully connected layer for classification. The model is trained using the cross-entropy loss function and the Adam optimizer with a learning rate of 0.0008. Training proceeds for 100 epochs with a batch size of 32, and the model weights yielding the highest validation accuracy are retained for final evaluation on the test dataset.
To enhance the capture of long-range temporal dependencies and refine feature representation, we further introduce an augmented architecture—denoted as InceptionTime CNN-Transformer—by integrating a Transformer encoder after the Inception modules. The feature maps extracted by the six Inception modules are reshaped and augmented with sinusoidal positional encodings, which inject order information using sine and cosine functions of varying frequencies. The resulting sequence is processed by a two-layer Transformer encoder employing a four-head self-attention mechanism. The output is then pooled globally and passed to the classification layer. The training protocol for this InceptionTime CNN-Transformer hybrid model remains identical to that of the standard InceptionTime CNN described above

\textbf{2D-CNN}: We design a CNN architecture employing two-dimensional convolutional layers, hereinafter referred to as the 2D-CNN model. The network accepts as input a visual encoding of orbits: each trajectory, with a length of 900 points sampled from the phase space $[0,1] \times [0,1]$, is encoded as a $64 \times 64$ pixel scatter plot, where the orbit is uniquely colored according to its pre-assigned class label.
This transformation allows the model to leverage spatial feature hierarchies inherent in the visual representation rather than raw coordinate sequences. The model is structured into a feature extraction module followed by a classification module. The feature extractor consists of three convolutional blocks. Each block contains a convolutional layer with a $3\times3$ kernel, stride 1, followed by a ReLU activation function and a $2\times2$ max-pooling operation. The number of channels increases across the blocks---32 in the first, 64 in the second, and 128 in the third. The resulting feature maps are then flattened into a one-dimensional vector. The classification module comprises two fully connected layers, each with 512 neurons and followed by ReLU activation and dropout regularization with a rate of 0.5. The final output layer produces logits for the four classes, which are passed through a softmax function to generate class probabilities. The model is trained using the Adam optimizer with a learning rate of 0.0008 and the cross-entropy loss function. Training employs mini-batches of 32 for up to 100 epochs, with early stopping implemented at a patience of 5. The model achieving the highest validation accuracy is saved and evaluated on the test dataset to assess classification performance.

\subsection{Deep Learning classification results}
\label{sec3.2}

Under the experimental configurations detailed for the aforementioned five models, we evaluate the classification performance of each model with varying parameters \(K\) and \(M\)---consistent with the parameters outlined in Figure \ref{fig1}---using identical training and testing datasets. The results, summarized in Table \ref{table1}, reveal that the 2D-CNN model demonstrates exceptional performance, consistently achieving classification accuracy exceeding 99\%. In contrast, the other four models exhibit significant performance fluctuations and failed to reach the same level of high, stable accuracy. The substantial performance gap suggests that the 2D-CNN is inherently well-suited for capturing spatially local patterns, which is particularly advantageous for this classification task.

\begin{table}[h]
	\centering
	\scalebox{0.7}{
		\begin{tabular}{|c|c|c|c|}
			\hline
			\diagbox{\textbf{Models}}{\textbf{Parameters}} & \textbf{K=0.01,M=3} & \textbf{K=0.03,M=3} & \textbf{K=0.1,M=3} \\
			\hline
			\textbf{MLP} & 80.50\%$\pm$0.50\% & 82.50\%$\pm$1.50\% & 90.50\%$\pm$0.50\% \\
			\hline
			\textbf{\tabincell{c}{MLP- \\ Transformer}} & 88.00\%$\pm$0.50\% & 84.00\%$\pm$0.50\% & 94.00\%$\pm$0.50\% \\
			\hline
			\textbf{\tabincell{c}{InceptionTime \\ CNN}} & 91.00\%$\pm$0.50\% & 90.00\%$\pm$0.30\% & 92.25\%$\pm$0.25\% \\
			\hline
			\textbf{\tabincell{c}{InceptionTime \\ CNN-Transformer}} & 92.00\%$\pm$0.50\% & 91.00\%$\pm$0.25\% & 94.75\%$\pm$0.25\% \\
			\hline
			\textbf{2D-CNN} & \textbf{99.50\%$\pm$0.25\%} & \textbf{99.45\%$\pm$0.45\%} & \textbf{99.50\%$\pm$0.25\%} \\
			\hline
			
		\end{tabular}
	}
	\vspace{1em}
	
	\scalebox{0.7}{
		\begin{tabular}{|c|c|c|c|}
			\hline
			\diagbox{\textbf{Models}}{\textbf{Parameters}} & \textbf{K=0.01,M=5} & \textbf{K=0.03,M=5} & \textbf{K=0.1,M=5} \\
			\hline
			\textbf{MLP} & 84.50\%$\pm$0.50\% & 91.00\%$\pm$1.25\% & 97.00\%$\pm$1.00\% \\
			\hline
			\textbf{\tabincell{c}{MLP- \\ Transformer}} & 86.25\%$\pm$0.50\% & 94.00\%$\pm$0.75\% & 97.50\%$\pm$0.50\% \\
			\hline
			\textbf{InceptionTime-CNN} & 91.50\%$\pm$0.25\% & 94.00\%$\pm$0.50\% & 97.25\%$\pm$0.25\% \\
			\hline
			\textbf{\tabincell{c}{InceptionTime \\ CNN-Transformer}} & 92.00\%$\pm$0.75\% & 96.50\%$\pm$0.75\% & 97.50\%$\pm$0.50\% \\
			\hline
			\textbf{2D-CNN} & \textbf{99.50\%$\pm$0.50\%} & \textbf{99.50\%$\pm$0.50\%} & \textbf{99.38\%$\pm$0.31\%} \\
			\hline
		\end{tabular}
	}
	\caption{Comparison of classification accuracy across the five DL models with different parameters.}
	\label{table1}
\end{table}

\begin{figure}[h]
	\centering
	\begin{subfigure}[b]{0.45\linewidth}
		\centering
		\includegraphics[width=0.9\linewidth]{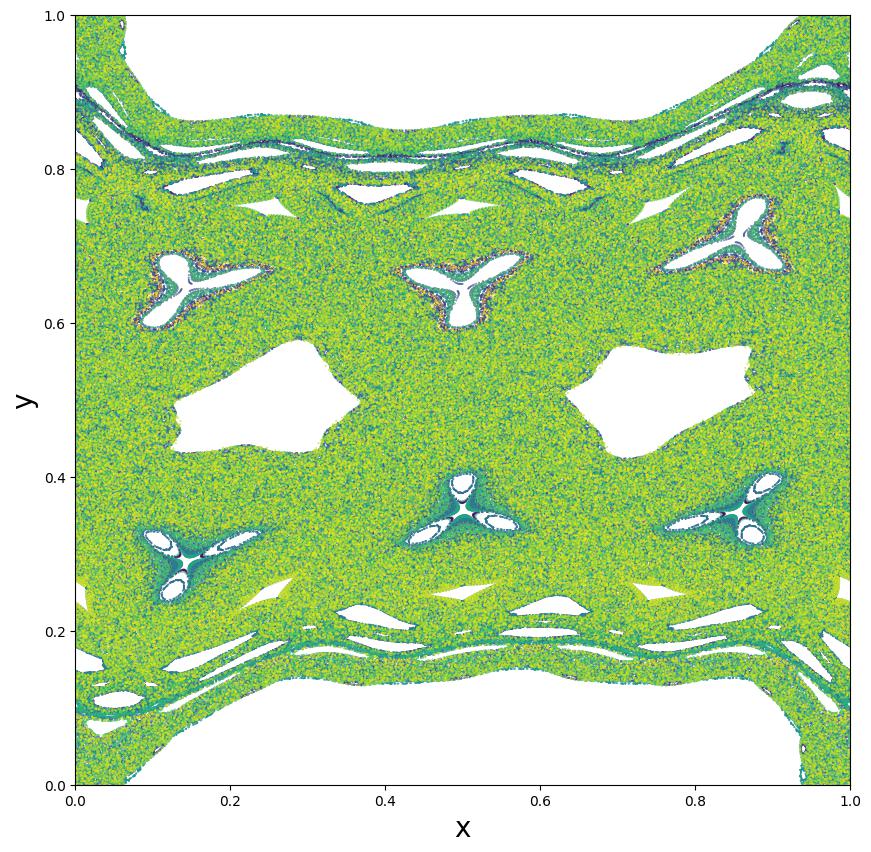}
		\caption{Strongly chaotic}
		\label{fig15a}
	\end{subfigure}
	\hfill
	\begin{subfigure}[b]{0.45\linewidth}
		\centering
		\includegraphics[width=\linewidth]{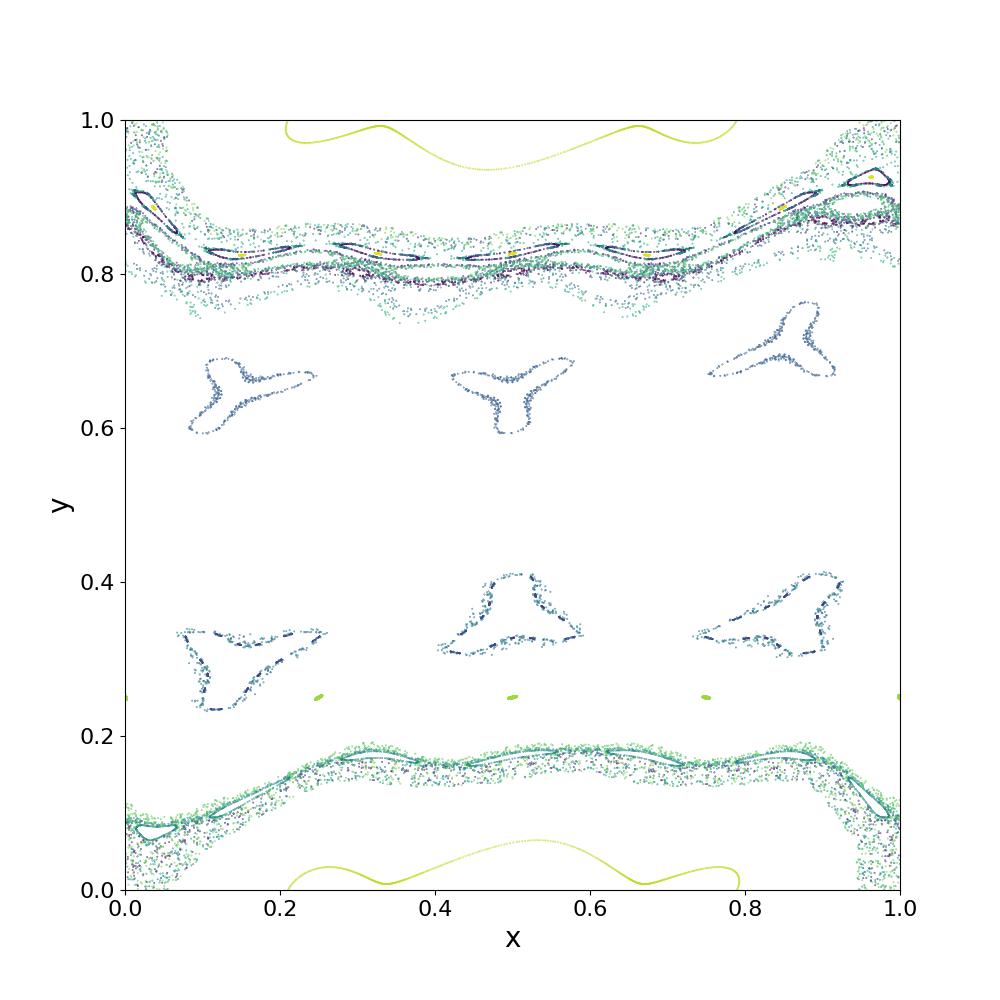}
		\caption{Weakly chaotic}
		\label{fig15b}
	\end{subfigure}
	\vspace{1em}
	\begin{subfigure}[b]{0.45\linewidth}
		\centering
		\includegraphics[width=\linewidth]{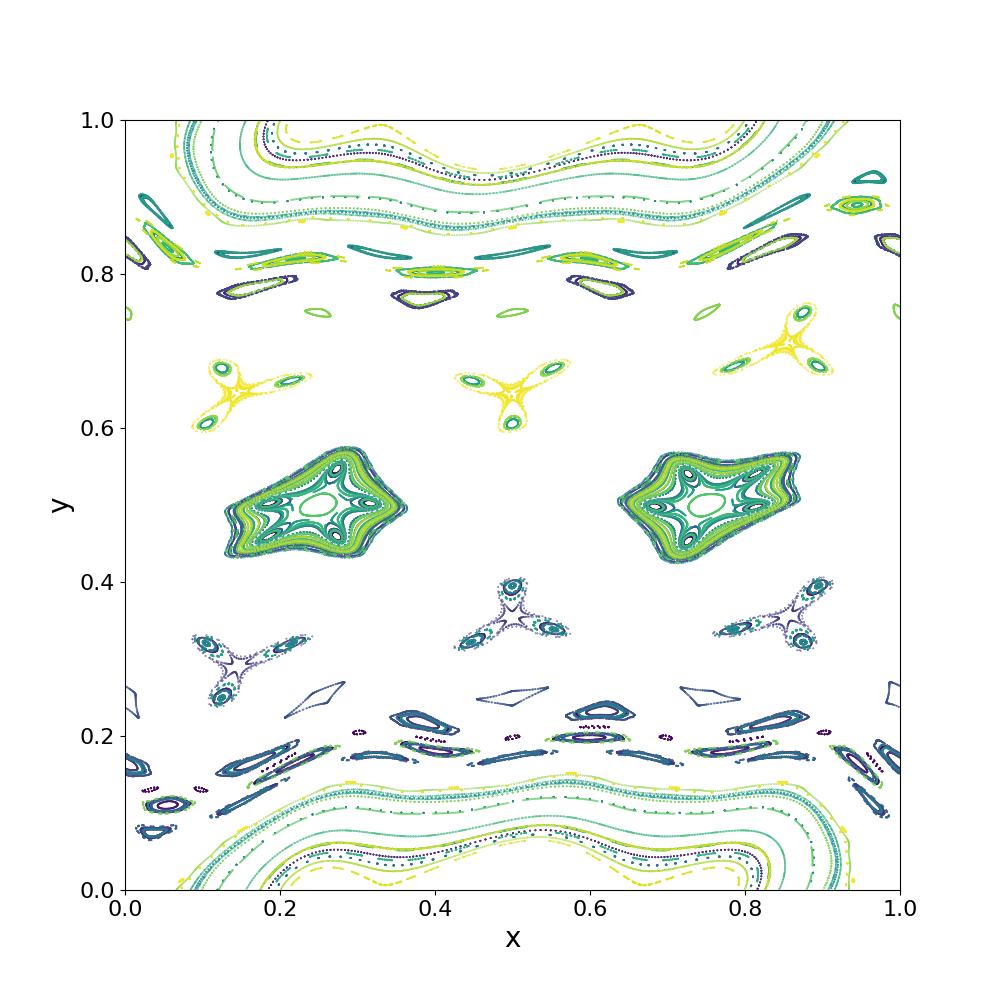}
		\caption{Resonant orbits}
		\label{fig15c}
	\end{subfigure}
	\hfill
	\begin{subfigure}[b]{0.45\linewidth}
		\centering
		\includegraphics[width=\linewidth]{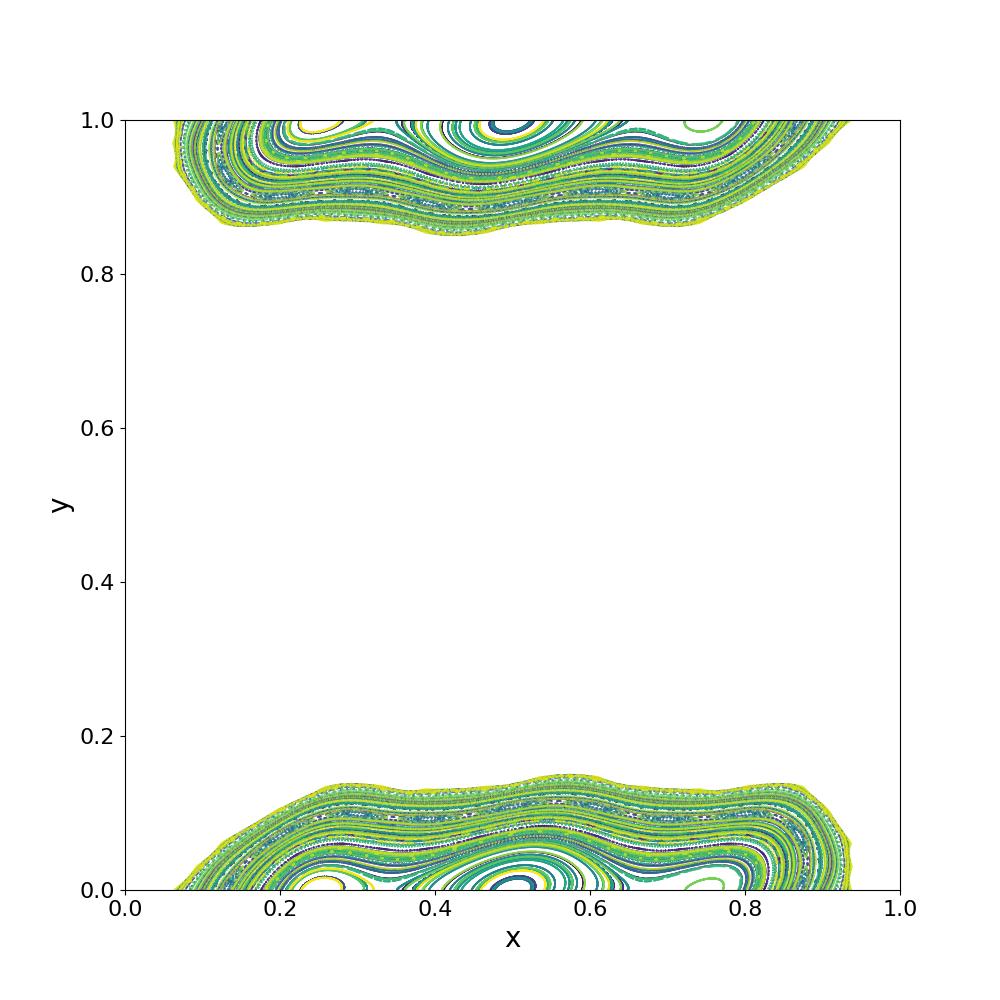}
		\caption{Non-resonant orbits}
		\label{fig15d}
	\end{subfigure}
	\caption{Trajectory diagrams of the four orbit types classified by the 2D-CNN model.}
	\label{fig15}
\end{figure}

To further validate the predictive reliability of the 2D-CNN model, we visualize trajectory diagrams for the four orbit types using the model’s predicted labels, as shown in Figure \ref{fig15}. These predicted trajectories exhibit near-perfect agreement with the trajectories displayed in Figures \ref{fig13g}, \ref{fig13h}, \ref{fig14b}, and \ref{fig14c}. This close visual correspondence demonstrates that the 2D-CNN model accurately captures the underlying dynamical structures, affirming its high fidelity in identifying the essential features of different orbital types.

\begin{figure}[h]
	\centering
	\begin{subfigure}[b]{0.45\linewidth}
		\centering
		\includegraphics[width=1.02\linewidth]{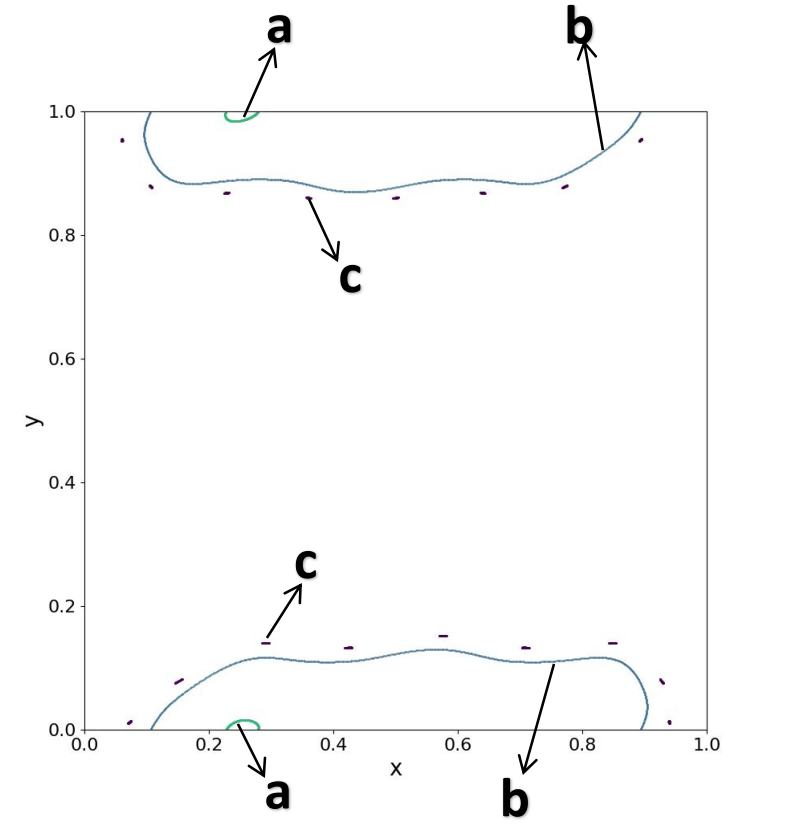}
		\caption{}
		\label{fig16a}
	\end{subfigure}
	\hfill
	\begin{subfigure}[b]{0.45\linewidth}
		\centering
		\includegraphics[width=\linewidth]{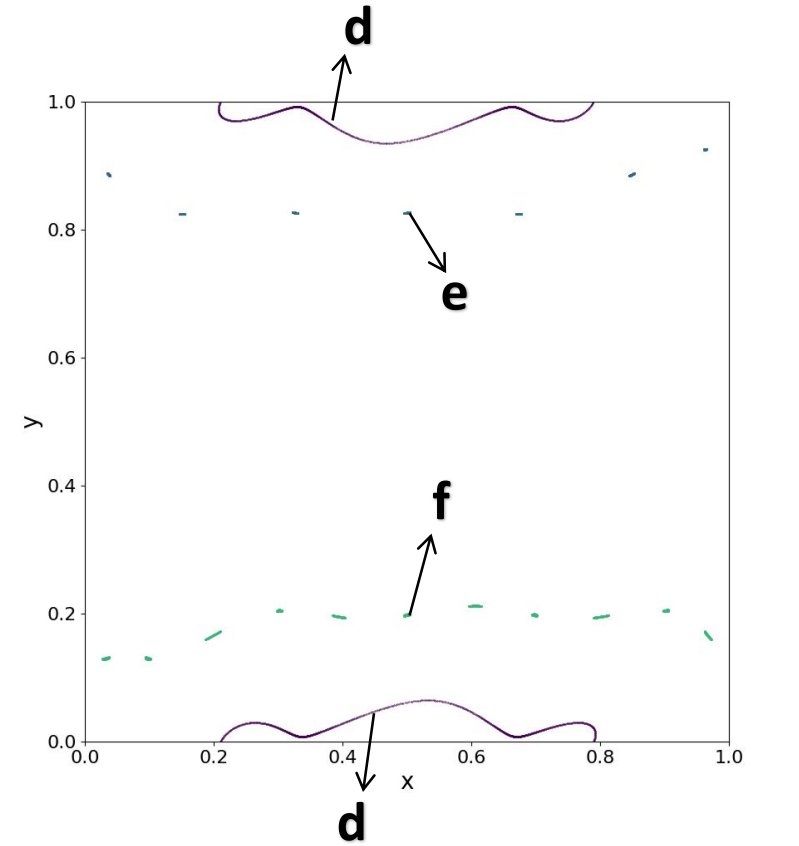}
		\caption{}
		\label{fig16b}
	\end{subfigure}
	\caption{(a)The numerical algorithm-based method misclassified the orbit, whereas the 2D-CNN accurately determined its type. (b)The numerical algorithm-based method correctly classified the orbit, whereas the 2D-CNN misidentified the orbit type.}
	\label{fig16}
\end{figure}

Although the dataset is largely reliable, it contains a small proportion of mislabeled orbits. The 2D-CNN model successfully reclassifies these erroneous cases into their appropriate dynamical categories. Table \ref{table2} provides a comparative summary of orbit-type distributions and misclassification counts for the numerical algorithm and the deep learning model. For instance, the numerical algorithm initially identified 137 orbits; however, validation using trajectory plots generated with longer iterations confirmed that 2 of these were misclassified. In contrast, the 2D-CNN model classified 133 orbits as resonant, with only one misclassification. To illustrate the corrective capability of the 2D-CNN, Figure \ref{fig16a} displays specific instances where orbits initially misclassified by the numerical algorithm were correctly identified by the deep learning model. For example, Orbit \textbf{b} was originally labeled as a resonant orbit by the numerical algorithm but was accurately reclassified as a non-resonant orbit by the 2D-CNN. In contrast, Figure \ref{fig16b} presents cases where the 2D-CNN misclassified orbits that were originally correct in the algorithmic labeling. For instance, Orbit \textbf{e}, correctly categorized as a resonant orbit by the numerical algorithm, was erroneously classified as a weakly chaotic orbit by the 2D-CNN. This phenomenon underscores that the deep learning model can capture fundamental patterns even when reference labels are wrong. Nevertheless, high-quality labeled data remains indispensable for reliably distinguishing challenging borderline cases, which are particularly susceptible to misclassification.

\begin{table}[h]
	\centering
	\scalebox{0.9}{
	\begin{tabular}{|c|c|c|c|c|}
		\hline
		\textbf{} & \textbf{\tabincell{c}{Numerical \\ algorithm}} & \textbf{\tabincell{c}{Classified \\ incorrectly}} & \textbf{2D-CNN} & \textbf{\tabincell{c}{Classified \\ incorrectly}} \\
		\hline
		\textbf{Class `0'} & 669 & 0 & 671 & 2\\
		\hline
		\textbf{Class `1'} & 137 & 2 & 133 & 1\\
		\hline
		\textbf{Class `2'} & 177 & 1 & 177 & 0\\
		\hline
		\textbf{Class `3'} & 17 & 0 & 20 & 3\\
		\hline
		\textbf{Total} & 1000 & 3 & 1000 & 6\\
		\hline
	\end{tabular}
}
	\caption{A comparative overview of the orbit counts for each dynamical type, obtained through the numerical algorithm and predicted by the trained 2D-CNN.}
	\label{table2}
\end{table}

\begin{figure}[h]
	\centering
	\begin{subfigure}[b]{0.3\linewidth}
		\centering
		\includegraphics[width=1.15\linewidth]{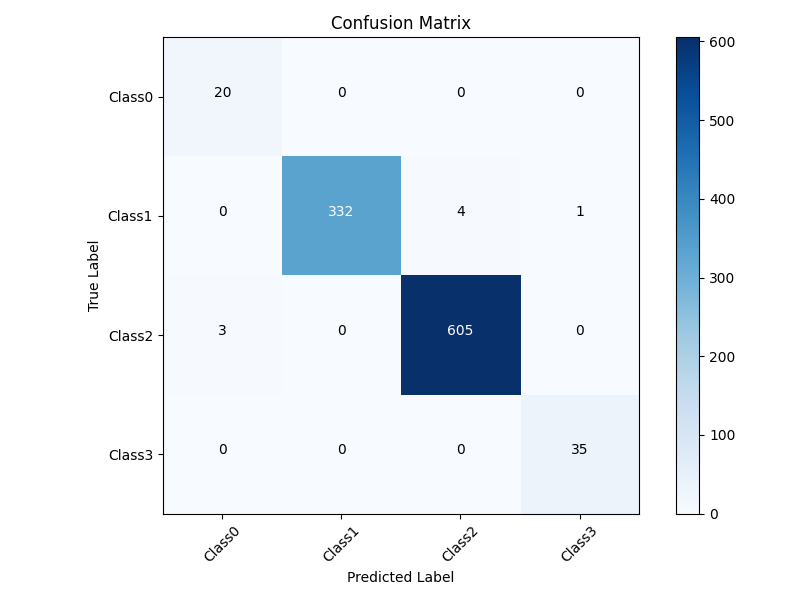}
		\caption{$K=0.008,M=3$}
		\label{fig17a}
	\end{subfigure}
	\hfill
	\begin{subfigure}[b]{0.3\linewidth}
		\centering
		\includegraphics[width=1.15\linewidth]{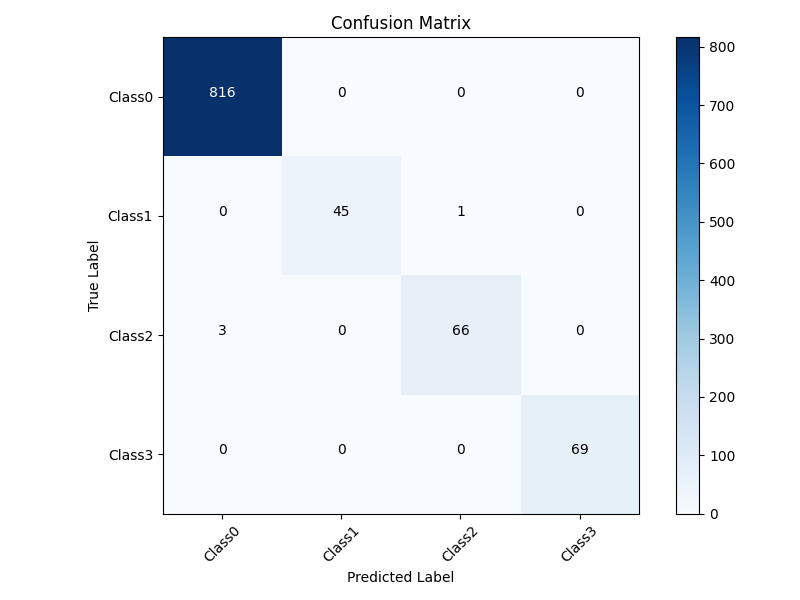}
		\caption{$K=0.024,M=5$}
		\label{fig17b}
	\end{subfigure}
	\hfill
	\begin{subfigure}[b]{0.3\linewidth}
		\centering
		\includegraphics[width=1.15\linewidth]{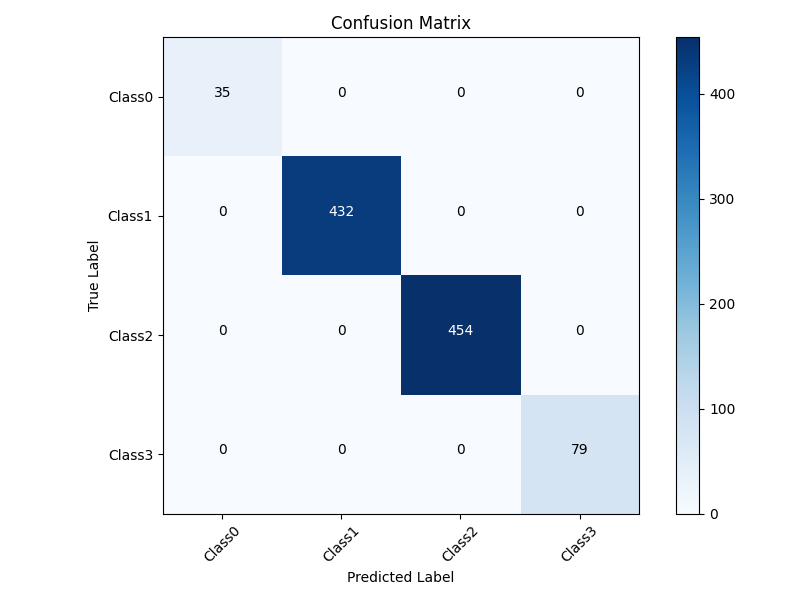}
		\caption{$K=0.032,M=2$}
		\label{fig17c}
	\end{subfigure}
	\vspace{0.2cm}
	\begin{subfigure}[b]{0.3\linewidth}
		\centering
		\includegraphics[width=1.15\linewidth]{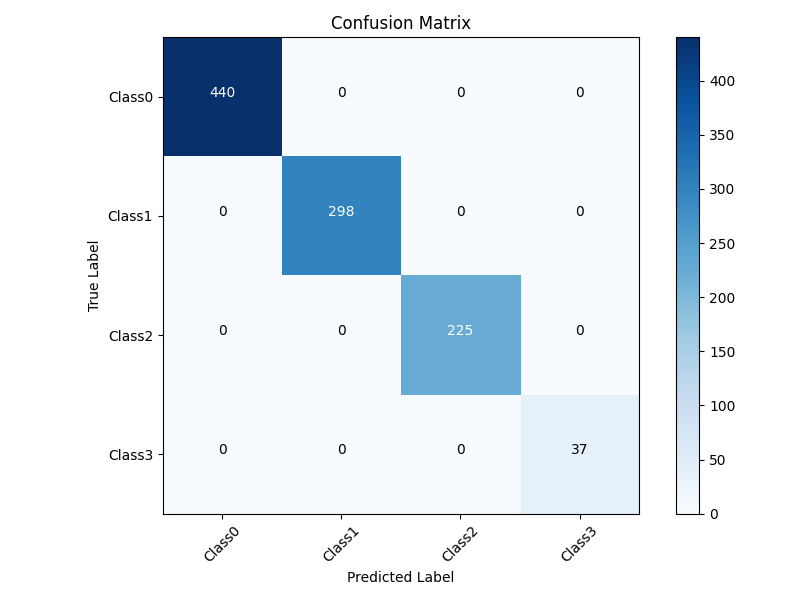}
		\caption{$K=0.054,M=2$}
		\label{fig17d}
	\end{subfigure}
	\hfill
	\begin{subfigure}[b]{0.3\linewidth}
		\centering
		\includegraphics[width=1.2\linewidth]{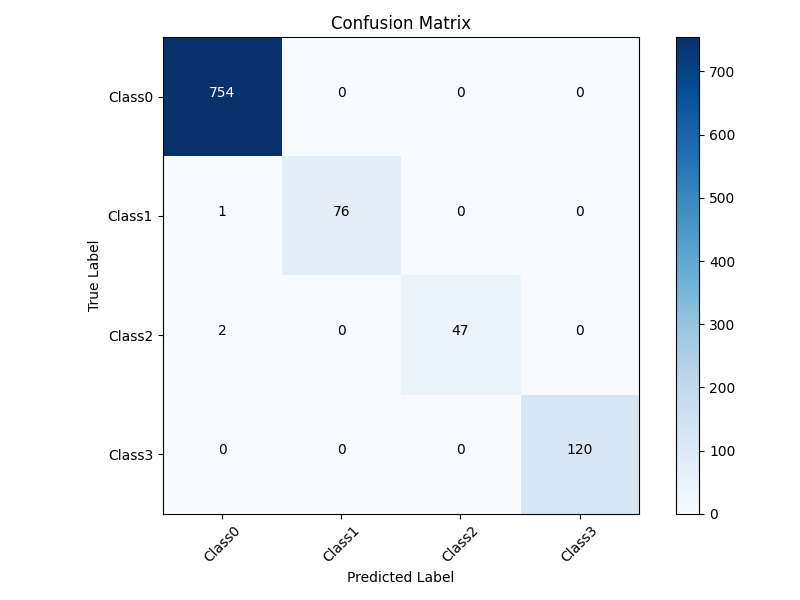}
		\caption{$K=0.06,M=3$}
		\label{fig17e}
	\end{subfigure}
	\hfill
	\begin{subfigure}[b]{0.3\linewidth}
		\centering
		\includegraphics[width=1.15\linewidth]{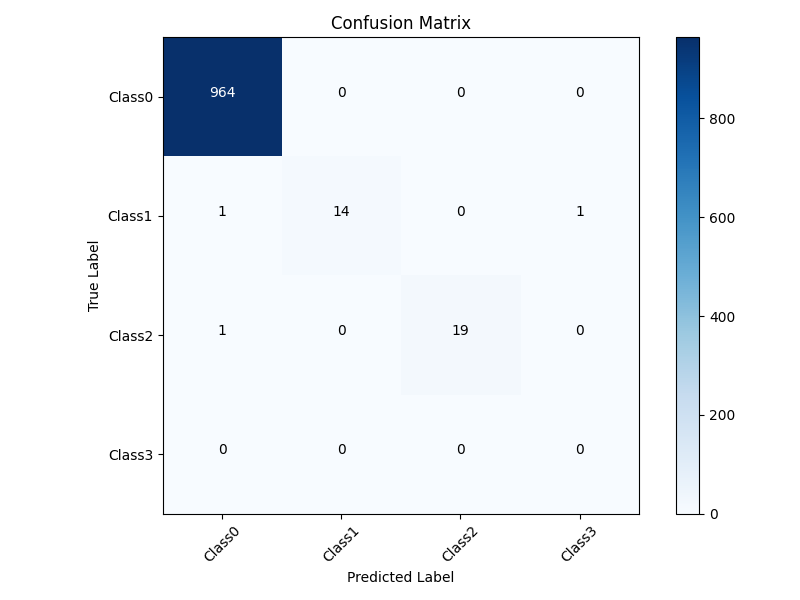}
		\caption{$K=0.09,M=3$}
		\label{fig17f}
	\end{subfigure}
	\caption{Classification performance of the 2D-CNN model evaluated under six distinct parameter pairs.}
	\label{fig17}
\end{figure}

\begin{figure}[t]
	\centering
	\begin{subfigure}[b]{0.3\linewidth}
		\centering
		\includegraphics[width=1.15\linewidth]{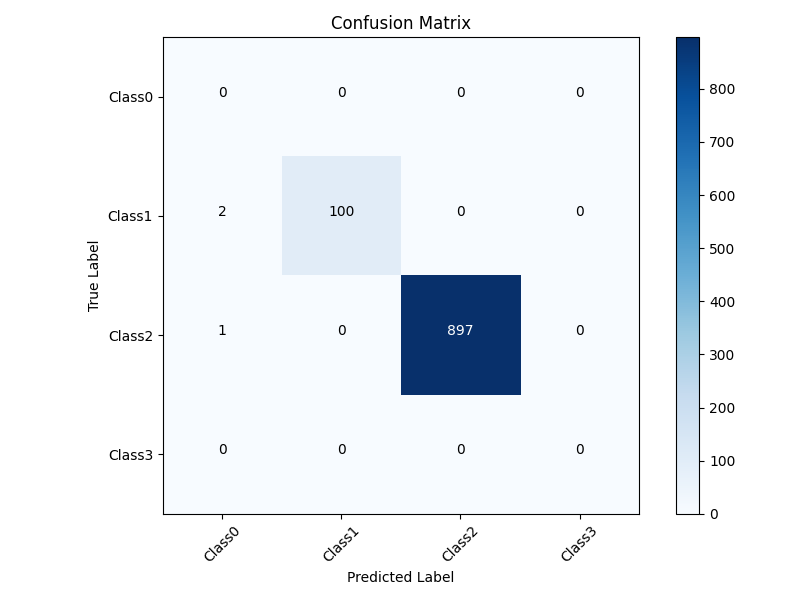}
		\caption{$K = 0.001, M = 3$}
		\label{fig18a}
	\end{subfigure}
	\hfill
	\begin{subfigure}[b]{0.3\linewidth}
		\centering
		\includegraphics[width=1.15\linewidth]{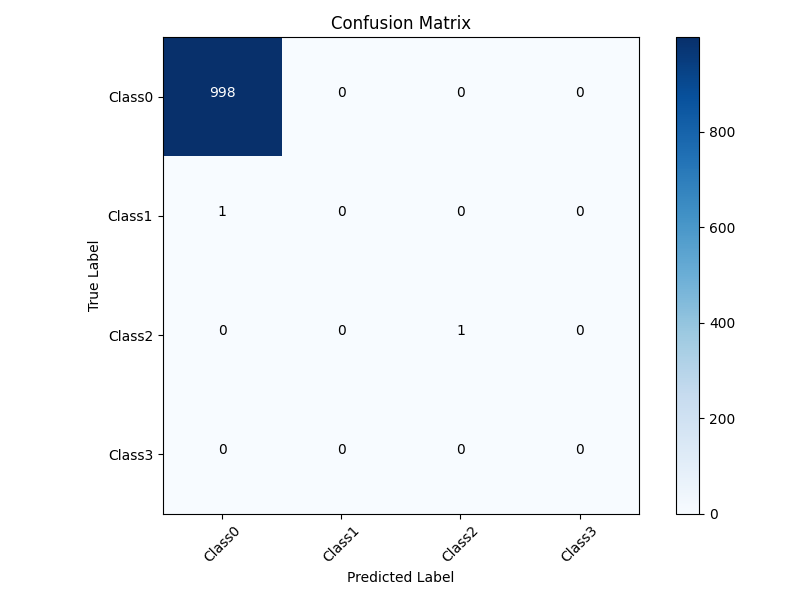}
		\caption{$K = 0.03, M = 10$}
		\label{fig18b}
	\end{subfigure}
	\hfill
	\begin{subfigure}[b]{0.3\linewidth}
		\centering
		\includegraphics[width=1.15\linewidth]{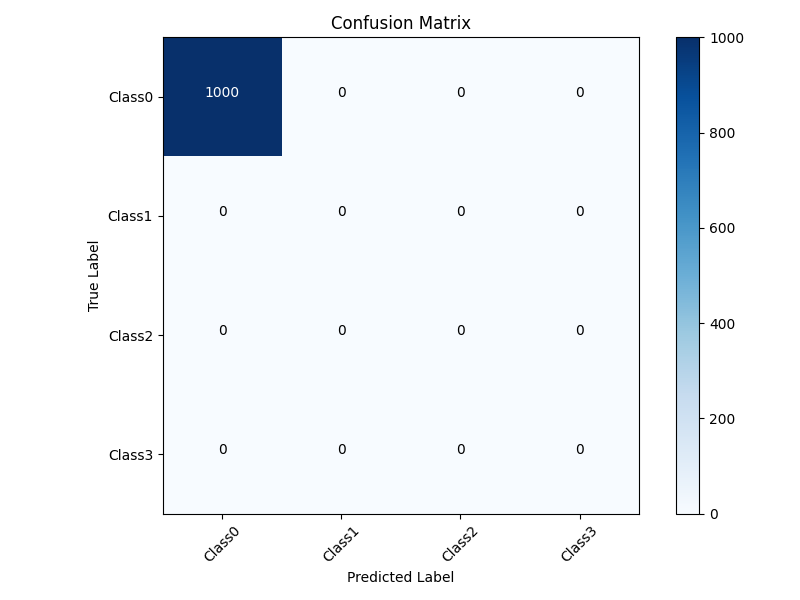}
		\caption{$K = 0.5, M = 3$}
		\label{fig18c}
	\end{subfigure}
	\caption{Generalization performance of the 2D-CNN on parameters outside the training domain.}
	\label{fig18}
\end{figure}

To systematically evaluate the classification performance of our 2D-CNN model across a continuous parameter space rather than at isolated points, we expanded our analysis to the intervals $K \in [0.005, 0.3]$ and $M \in [2, 6]$. For each integer value of $M$ in this range, ten values of $K$ were randomly sampled, resulting in a total of 50 distinct parameter pairs. For every resulting parameter pair $(K, M)$, we first generated the dynamical labels using our numerical algorithm. These labels were then paired with the corresponding 900-length orbital segments to form the individual data subsets. Finally, all subsets from the different parameter pairs were combined into a single dataset for deep learning. To assess the model's performance on unseen parameter combinations within the same operational domain, we selected six distinct pairs from these intervals and evaluated each in a separate test run, ensuring no overlap with the training pairs. As illustrated in Figure \ref{fig17}, the model achieved exceptional classification accuracy, surpassing 99\%. Moreover, as shown in Figure \ref{fig18}, it maintained high performance when evaluated on parameter pairs outside the training intervals (specifically, $K = 0.03, M = 10$; $K = 0.5, M = 3$; and $K = 0.001, M = 3$), confirming its robust generalization capability. These results indicate that the model captures underlying dynamical patterns consistent across the parameter space, rather than overfitting to specific training examples.

\section{Conclusion}
\label{sec4}

Our study presents a novel framework for the granular classification of orbital dynamics within the GKR system, successfully distinguishing four distinct categories: strongly chaotic, weakly chaotic, resonant, and non-resonant orbits. A key innovation lies in the synergistic integration of three complementary methods---the weighted Birkhoff average, Lyapunov exponent, and correlation dimension---which enables robust identification, particularly of the elusive weak chaos regime where single-metric approaches often fail. This foundational work facilitated the creation of a high-quality, algorithmically-labeled dataset pairing long-term integration labels with shorter trajectory segments, effectively balancing classification accuracy with practical applicability for deep learning.

Subsequently, we demonstrated that a well-trained 2D-CNN architecture excels in this classification task, significantly outperforming models including MLP, InceptionTime CNN, and their hybrid Transformer variants (MLP-Transformer and InceptionTime CNN-Transformer). The superior performance is attributed to the network's ability to capture the inherent spatial patterns in the trajectory data. Moreover, the 2D-CNN exhibited a corrective capability, refining the initial algorithmic labels by rectifying some misclassifications, which underscores its ability to capture underlying dynamical principles beyond the training labels. Additionally, the model demonstrates considerable generalization capability across varied dynamical regimes, confirming its robustness. To our knowledge, this work represents the first successful application of machine learning to classify a dynamical system into four distinct orbital categories, providing an intuitive visualization that enhances the understanding of weakly chaotic orbits.

\section{Acknowledge}
This work was supported by the National Natural Science Foundation of China (Grant No. 12271400 to Z. Xu) and the Science and Technology Development Plan Project of Jilin Province (Grant No. 20250102016JC to J. Zu).

\appendix
\section{Numerical identification of rational numbers}
\label{Appendix A}

\begin{figure}[h]
	\centering
	\begin{subfigure}[b]{0.45\linewidth}
		\centering
		\includegraphics[width=1.2\linewidth]{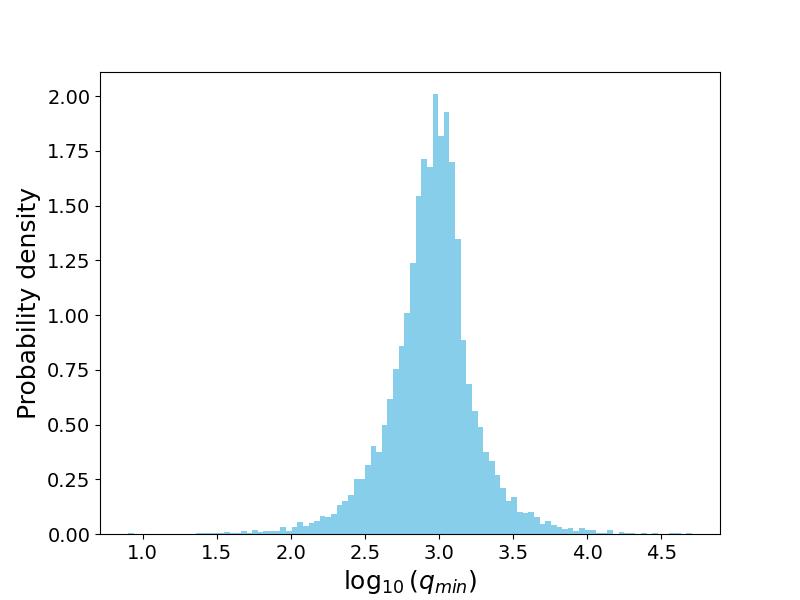}
		\caption{}
		\label{fig19a}
	\end{subfigure}
	\hfill
	\begin{subfigure}[b]{0.45\linewidth}
		\centering
		\includegraphics[width=1.05\linewidth]{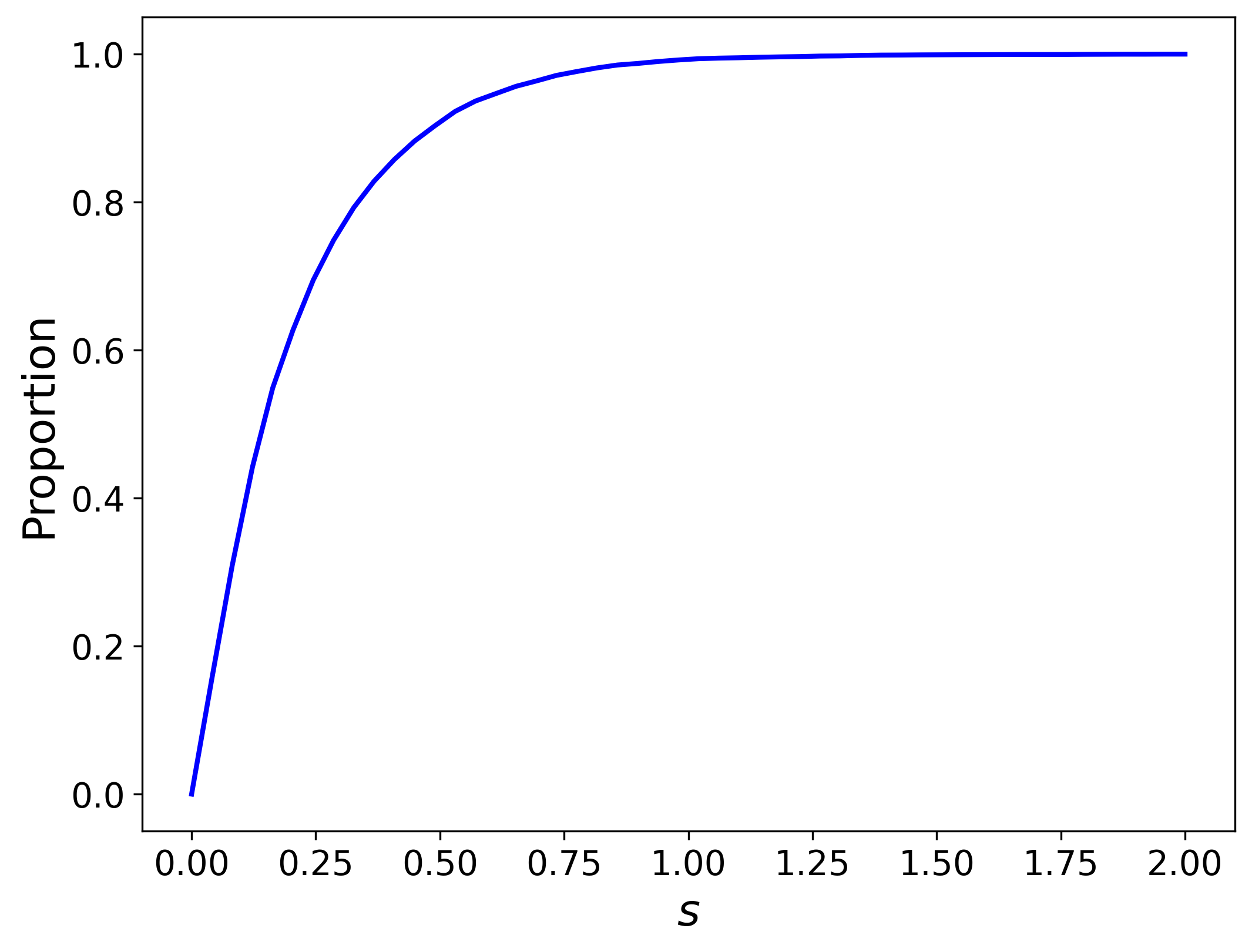}
		\caption{}
		\label{fig19b}
	\end{subfigure}
	\caption{(a) Probability density of $\log_{10}(q_{\text{min}})$ with $\delta = 10^{-6}$ for $10^4$ randomly chosen floating-point numbers in $[0, 1]$. (b) A graph of the probability that $q_{\min} \in [10^{3-s}, 10^{3+s}]$.}
	\label{fig19}
\end{figure}

Given a rotation number $\omega \in [0,1]$, and an interval $I_\delta(\omega)$ for a small $\delta$:
\[
I_\delta(\omega) \equiv \left(\omega-\delta, \omega+\delta \right). \qquad \delta=10^{-tol}
\]
We denote the smallest denominator for a rational in an interval $I_\delta(\omega)$ by:
\[
q_{\min}(I_\delta) \equiv \min \left\{ q \in \mathbb{N} : \frac{p}{q} \in I_\delta,\ p \in \mathbb{Z} \right\}.
\]
\cite[Algorithm 2]{sander2020birkhoff} provides a numerical method for finding $q_{\text{min}}$. Applying this algorithm to $10^4$ randomly chosen floating-point numbers in $(0, 1)$ with uniform distribution, we compute $q_{\text{min}}$ and plot the histogram of $\log_{10}(q_{\text{min}})$(see Figure \ref{fig19a}).

We know that if there exists a rational number $p/q \in I_\delta(\omega)$ with a sufficiently small denominator $q$, then $\omega$ is well-approximated by this rational. Conversely, if all such rationals within $I_\delta(\omega)$ have large denominators, we would typically expect $\omega$ to approximate an irrational number. However, if the denominator $q$ is excessively large, it may indicate that $\omega$ is more likely an approximation of a rational number that narrowly missed being within the interval. To quantify whether $q$ is small, large, or excessively large, \cite{sander2020birkhoff} introduces a quantitative indicator $\text{dev}_{\omega}$:
\[
\text{dev}_{\omega} = |\log_{10}(q_{\text{min}}( I_\delta(\omega)))-tol/2|
\]

We classify rotation numbers based on this indicator, if $\text{dev}_{\omega} > s$, then denominator $q$ is either small or excessively large, and we consider the rotation number to more closely approximate a rational. Otherwise, $\omega$ is deemed closer to an irrational. In this paper, we set $\delta=10^{-6}$ and $s = 0.67$, which implies 95.65\% of the randomly sampled values, $q_{\text{min}}$ fall within the interval $[10^{3-s}, 10^{3+s}]$, as shown in Figure \ref{fig19b}.

\bibliographystyle{elsarticle-num-names}
\bibliography{ref}

\end{document}